\documentclass[manuscript]{aastex} 
 
\begin{document} 
 
\title{Cloud Structure and Physical Conditions in Star-forming Regions 
       from Optical Observations. ~I. Data and Component Structure} 
 
\author{K. Pan,\altaffilmark{1,2,3} ~~ S. R. Federman,\altaffilmark{1,2}} 
\affil{Department of Physics and Astronomy, University of Toledo, 
    Toledo, OH 43606} 
     
\author{K. Cunha,} 
\affil{Observatorio Nacional - MCT, Rua Gal Jose Cristino 77, 
    Rio de Janeiro, Brazil CEP 22921-400} 
     
\author{V. V. Smith,} 
\affil{Department of Physics, University of Texas at El Paso, 
    El Paso, TX 79968\\ 
    and McDonald Observatory, University of Texas at Austin, Austin, TX 78712}     
\and 
 
\author{D. E. Welty\altaffilmark{1,2}} 
\affil{University of Chicago, Astronomy and Astrophysics Center, 5640 South  
Ellis Ave., Chicago, IL 60637} 
 
\altaffiltext{1}{Guest Observer, McDonald Observatory, University of Texas  
at Austin.}

\altaffiltext{2}{Visiting Observer, Kitt Peak National Observatory, National 
 Optical Astronomy Observatories, which is operated by AURA, Inc., under  
cooperative agreement with the National Science Foundation.} 
 
\altaffiltext{3}{Current address: Department of Physics and Astronomy, Bowling  
Green State University, Bowling Green, OH 43403} 
 
\begin{abstract} 
We present high-resolution optical spectra (at $\sim$0.6--1.8 km s$^{-1}$) of  
interstellar CN, CH, CH$^{+}$, \ion{Ca}{1}, \ion{K}{1}, and \ion{Ca}{2}  
absorption toward 29 lines of sight in three star-forming regions, $\rho$ Oph, 
Cep OB2, and Cep OB3. The observations and data reduction are described. 
The agreement between earlier measurements of the total equivalent widths and 
our results is quite good. However, our higher resolution spectra reveal  
complex structure and  
closely blended components in most lines of sight. The velocity component  
structure of each species is obtained by analyzing the spectra of the six  
species for a given sight line together. The tabulated column densities 
and Doppler parameters of individual components are determined by using 
the method of profile fitting. Total column densities along lines 
of sight are computed by summing results from profile fitting 
for individual components and are compared with column densities from the 
apparent optical depth method. A more detailed analysis of  
these data and their implications will be presented in a companion paper. 
\end{abstract} 
 

\section{INTRODUCTION} 
The interstellar medium (ISM) is not static as was thought during the early  
part of 20$^{th}$  
century. It is the reservior out of which stars are born and into which  
stars inject newly created elements as they age. The physical and chemical  
state in the ISM continuously changes by the combined actions of star  
formation and stellar death (Cox \& Smith 1974; Jenkins \& Meloy 1974;  
Burstein 1977). The evolution of a galaxy is governed to a large extent by  
these processes. The final stages of star formation involve the disruption  
of interstellar material from which stars form. If enough material is  
available in interstellar clouds forming O and B stars, one generation of  
stars can initiate the formation of a second generation, and so forth, but  
eventually the cloud remnants will no longer be able to sustain star 
formation. The propagation of star formation occurs in regions of enhanced  
gas density behind shocks (Elmegreen \& Lada 1977). Phenomena associated  
with O and B stars that lead to shocks in the surrounding medium include  
stellar winds, expanding H~{\small II} regions, and supernova explosions.  
Since O and B stars tend to form in clusters, these effects are magnified;  
the prodigious star formation seen in the Magellanic Clouds is an example  
of these processes. Clearly, knowledge of the physical conditions and  
chemical composition of the ISM in star-forming regions will help us  
understand the above processes, in particular, and the evolution of the  
ISM and the Galaxy in general.  
 
Determination of the physical properties and chemical transitions of  
individual interstellar clouds from observations of absorption lines  
requires high-resolution spectra. High resolution is needed to distinguish  
the individual components contributing to the generally complex  
absorption-line profiles seen in most lines of sight. Most previous studies  
utilized one of two complementary approaches. 
One approach relies on a comprehensive analysis of the atomic and molecular 
material observed along a specific line of sight toward a background star.  
The objective here is the determination of cloud structure, density, 
and temperature for the gas and of the flux of ultraviolet radiation  
permeating the cloud. Efforts in this vein include analyses of gas toward  
$\zeta$ Oph (Black \& Dalgarno 1977; van Dishoeck \& Black 1986),  
$\gamma$ Ara (Federman \& Glassgold 1980),  
HD 169454 (Januzzi et al. 1988), HD 29647 and HD 147889 (van Dishoeck \& Black  
1989), and HD 37903 and HD 147009 (Knauth et al. 2001). The second approach  
focuses on a single species through observations of many lines of sight. 
Previous studies of this kind examined the chemistry of CO (Federman et  
al. 1980), of CH (Federman 1982; Danks, Federman, \& Lambert 1984;  
Crawford 1989), and of CN (Federman, Danks, \& Lambert 1984). The main goals  
of these investigations are to study the 
chemistry of observed species and to put constraints on chemical and  
physical conditions in clouds.  
 
  In the present study, we combine the two approaches. Interstellar  
material toward 29 lines of sight in three star forming regions, $\rho$ Oph,  
Cep OB2, and Cep OB3, is probed through a study of absorption lines against  
the continua of background stars. High-resolution optical spectra  
($\delta$v $\sim$ 0.6--1.8 km s$^{-1}$) of six species, CN, CH, CH$^{+}$,  
\ion{Ca}{1}, \ion{K}{1}, and \ion{Ca}{2} are obtained for each line of sight.  
The present study provides the most complete picture of the physical  
conditions and cloud structure in the diffuse molecular gas associated with 
the three star-forming  
regions. While many observations of radio emission from these  
three regions have been made, the much better sensitivity of absorption  
line studies allows us to observe gas components with lower molecular 
column density  
that radio observations cannot find and to reveal small-scale cloud  
structure that radio observations did not uncover. On the other hand,  
comparison of the  
absorption line measurements to the large scale CO radio observations can  
help us understand the large-scale structure of clouds.  
 
In this paper, we present the optical spectra for the three star-forming  
regions. In $\S$$\S$ 2 and 3, we describe the  
procedures to obtain, reduce, and analyze the spectra. A more detailed  
analysis of these spectra and their implications will be presented in a  
companion paper (Pan et al. 2003).  
 
\section{OBSERVATIONS AND DATA REDUCTION} 
 
\subsection{Program Stars} 
 
     Observations of absorption from interstellar gas require the use of 
a bright, ideally, rapidly rotating star, as the light source; sharp  
interstellar absorption lines are then superposed on broader stellar features. 
Typically, there are multiple interstellar clouds (or cloud clumps) along 
a given line of sight whose existence is revealed by absorption at distinct 
Doppler velocities.  In order to study 
cloud structure and physical conditions of the clouds associated with 
$\rho$ Oph, Cep OB2, and Cep OB3, interstellar CN, CH, CH$^{+}$, \ion{Ca}{1},  
\ion{K}{1}, and \ion{Ca}{2} absorption was sought in 29 sight lines.  
 
Our target list is based on earlier stellar and interstellar studies 
(e.g., Federman et al. 1994; Patel et al. 1998; Daflon et al. 2001).  
Stellar data for the program stars are listed  
in Table 1. In particular, the HD number, name, spectral type, $V$, Galactic 
coordinates, reddening, and distance are given. 
The values of $E(B-V)$ are from Blaauw et al. (1959), Simonson (1968), and  
Garrison \& Kormendy (1976), and the distances are estimates from a 
combination of Hipparcos measurements (Perryman 1997) and spectroscopic 
parallax using data from Allen (1973). Other parameters are from the Simbad  
Database, operated at CDS, Strasbourg, France.  
An additional selection criterion for Cep OB2 was the position of  
the line of sight relative to cloudlets seen in CO radio emission (Patel et al. 
1995, 1998).  
   
Spectra were acquired at two sites. Brighter stars were observed with the  
0.9 m Coud\'{e} Feed Telescope at Kitt Peak National Observatory (KPNO),  
while fainter stars were observed with the 2.7 m telescope at McDonald  
Observatory of the University of Texas at Austin. Data on three stars,  
$\rho$ Oph A, HD203374A, and HD206773, were obtained at both sites,  
thereby providing a check on consistency.   
Additional observations of stars in Cep OB2 were obtained with the 
Sandiford echelle spectrograph on the 2.1m telescope at McDonald 
Observatory; details about data reduction can be found in 
Daflon et al. (2001). These spectra included the wavelength 
range sampling CH, CH$^+$, and \ion{Ca}{1} absorption.  Interstellar  
equivalent widths for sight lines from this data set,  
but not studied at high spectral resolution, are given in the Appendix.
 
\subsection{KPNO Spectra} 
     
Interstellar spectra for 12 program stars were obtained at KPNO during  
four runs in 1995 (September 28---October 5), 1996 (September 4---5),  
1999 (October 30---November 8), and 2000 (September 
9---12). Three instrumental setups were  
used in order to get CN, CH, CH$^{+}$, \ion{Ca}{1}, \ion{K}{1}, and  
\ion{Ca}{2} absorption features for these lines of sight.  
 
Two configurations were used for blue settings. The first setup was for  
interstellar CN $\lambda$3874 and \ion{Ca}{2} $\lambda$3933 absorption. To  
achieve the desired high spectral resolution ($\sim$ 1.5 km s$^{-1}$), we  
used  the echelle grating in 143$^{th}$ order centered near 3943 \AA, cross  
dispersed via grism 780-2 in 2$^{nd}$ order, camera 6,  and a 90 $\mu$m  
entrance slit, corresponding to 0.$^{\prime\prime}$6 on the sky. A CuSO$_4$  
filter was used to eliminate the grism's 1$^{st}$ order contributions, and a  
3$^{\circ}$ wedge positioned the desired spectral region on the F3KB CCD  
(1024$\times$3072, 15 $\mu$m pixels) detector. Exposures were binned by 2  
pixels perpendicular to the dispersion. For the second setup, interstellar  
absorption features of CH $\lambda$4300, CH$^{+}$ $\lambda$4232, and 
\ion{Ca}{1} $\lambda$4226 were obtained by centering the 132$^{nd}$ order  
of the echelle grating spectrum near 4273 \AA, keeping other settings the  
same as in the first setup. Grism 730 instead of 780-2 was 
employed, and no wedge was used for this setup in 1995.  
 
A third setup was used to acquire interstellar \ion{K}{1} $\lambda$7699  
features by centering the 73$^{rd}$ order of the echelle grating spectrum  
near 7700 \AA. In this setup,  an RG-610 blocking filter was 
used. The entrance slit width was changed to 100 $\mu$m. 
Other settings were the same as in the second setup. 
 
In addition to stellar images, a variety of exposures were made for the  
purpose of calibration. Fifteen zero and flat lamp exposures, and a few  
Th-Ar hollow cathode lamp exposures, were taken each night. Five dark  
exposures were made during each observing run. The zero exposures were used  
to remove the bias voltage in the CCD detector. 
Flat lamp frames measured the variation in  
sensitivity from pixel to pixel across the CCD chip. 
The Th-Ar comparison exposures provided the wavelength scale for the extracted  
interstellar spectra. Dark frames   
gave the maximum contribution from thermal noise, which was subtracted from  
stellar exposures when necessary. Moreover, the echelle  
grating was moved slightly each night in order to reduce the effect of any  
flat field artifacts. The setups yielded spectral resolutions of about 1.3  
to 1.5 km s$^{-1}$, as determined from the full widths at half maximum (FWHM)  
of the thorium lines in the comparison spectra assuming intrinsic widths of  
the lines are about 0.55 km s$^{-1}$ (Welty et al. 1994).  
 
\subsection{McDonald Observatory Spectra} 
   Twenty stars were observed with the 2dcoud\'{e} spectrograph (cs21) (Tull  
et al. 1995) on the 2.7 m telescope during 
three runs in 2000 (September 3---8)  
and 2001 (July 7---13; August 29---31). Grating E1 and the TK3 CCD  
were used for the observations. A cross-dispersed echelle spectrometer and  
the availabiltity of a 2048$\times$2048 CCD made it possible to get CN, CH,  
CH$^{+}$, \ion{Ca}{1}, and \ion{Ca}{2} absorption features in a single  
exposure by setting the cross disperser to its 
2$^{nd}$ order, and by centering  
the 56$^{th}$ order of the echelle grating spectrum to 4065 \AA. \ion{K}{1}  
$\lambda$7699 spectra were obtained by setting the cross disperser to its  
1$^{st}$ order and by centering the 31$^{st}$ order spectrum to 7165 \AA.  
Slit \# 2 with a width of 145 $\mu$m was used for all observations. An  
exposure time of 30 minutes per stellar image for faint stars was chosen to  
limit the effect of cosmic rays during the data collection process.   
 
As we did in KPNO observations, a variety of auxiliary exposures were taken 
for calibration purposes. Typically, four or five dark exposures were made  
for each observing run. Twenty zeros and ten flat lamp exposures were taken  
each night. Th-Ar comparison frames were taken throughout each night, usually  
every two hours. The measured width of thorium emission lines in the  
comparison spectra indicates a resolution of about 1.7 km s$^{-1}$ for these  
observations.  
 
Higher resolution spectra, of about 0.56 km s$^{-1}$, for \ion{K}{1} toward 
9 Cep, $\nu$ Cep, and $\lambda$ Cep, and for \ion{Ca}{1} toward $\lambda$ Cep  
were obtained with the same telescope and a double-pass configuration in  
1995 and 1996 (for details of the instrumental setup, see Welty \& Hobbs 2001; 
Welty, Hobbs, \& Morton 2003). These spectra are labeled as M2 in  
Figures 1---15. 
 
\subsection{Data Reduction}  
 
The raw image data from both telescopes were reduced in the usual manner  
with the NOAO IRAF (Image Reduction and Analysis Facility) package.  
Flat field frames were first examined to determine overscan and data regions,  
while object and bias frames were examined to locate abnormal (dead or hot) 
pixels. Stellar, flat, and 
comparison frames were bias-corrected by subtracting the 
average bias exposure from them. Cosmic-ray hits were removed from stellar 
images. Scattered light was fitted by low-order polynomials, in both the 
dispersion direction and perpendicular to it for the multi-order echelle  
data, and removed from each stellar and flat field exposure. The normalized 
average flat field was divided into the stellar images to account for  
differences in pixel-to-pixel sensitivity. Since each echelle order was a  
few pixels wide, the orders were summed across this width without weight.  
Then all extracted one-dimensional spectra  
were calibrated in wavelength using the Th-Ar comparison spectra, and were 
Doppler-corrected.  
Spectra with interstellar absorption lines and sufficient continua on both  
sides, usually 2 \AA, were cut from each stellar exposure. Each spectrum was  
carefully examined for flat field artifacts and cosmic ray blemishes that  
remained. The cosmic ray blemishes were subsquently  
removed unless the blemishes coincided with interstellar lines; then that  
spectrum was excluded in the final summation. After the  
examination, spectra for the same species toward the same star were summed  
together to obtain a final spectrum with higher signal-to-noise ratio, SNR,  
typically about 50 to 150. The stellar continua were normalized by fitting  
low order Legendre polynomials to regions free of interstellar absorption.  
This continuum normalization was easily performed except for  
a few cases, such as \ion{Ca}{2} toward $\nu$ Cep, 9 Cep, and 13 Cep where  
interstellar features were superposed on strong, relatively narrow stellar  
absorption lines. As a result, higher order polynomials had to be used there. 
The data reduction for K95, K96, M95, and M96 spectra is somewhat different 
(see Welty \& Hobbs 2001 for details).  
 
The data reduction procedure used here is slightly different from the 
one employed by Pan et al. (2001). Instead of median smoothing, the  
{\bf cosmicrays}  
routine was used to remove cosmic-ray hits, bad CCD columns were fixed by 
linear interpolation from neighboring columns, and all frames were overscan  
debiased prior to average-bias subtraction. The new procedure yielded more  
satifactory continua. The overall shapes of line profiles did not  
change, although line profiles were slightly deeper in some cases because 
the original smoothing broadened lines artificially. We were 
able to discover a few weak components that were not visible with the 
original procedure. However, Pan et al.'s 
(2001) conclusions are not altered by the improved extraction.  
       
The final normalized interstellar spectra for CN, CH, CH$^{+}$, \ion{Ca}{1},  
\ion{K}{1}, and \ion{Ca}{2} are shown in Figures 1 through 15.   
The spectra are shown in $V_{LSR}$ space.  
To facilitate comparisons between profiles, all the \ion{K}{1} and  
\ion{Ca}{2} spectra are on the same vertical scale, but the scales for weaker  
lines may differ from panel to panel.  
 
\section{ANALYSIS AND RESULTS} 
 
\subsection {Total Equivalent Widths} 
 
Some of our program stars were observed previously, usually at lower 
resolution; total equivalent widths ($W_{\lambda}$) along a line of  
sight were given. In order to compare our results with the previous values  
of $W_{\lambda}$, we also measured total equivalent widths for all six  
species. Tables 2---4 list our measurements along with their uncertainties  
and previous $W_{\lambda}$ values from the literature. 
Three stars, $\rho$ Oph A, 
HD203374A, and HD206773, were observed at both McDonald and Kitt Peak.  
Therefore, two measurements exist for all species toward them except for  
CH toward $\rho$ Oph A, for which we had only McDonald spectra, and for  
\ion{K}{1} toward HD203374A and HD206773,  
which were observed at Kitt Peak only. The first rows for the three stars are  
$W_{\lambda}$ values from McDonald spectra, and the second rows are 
from Kitt Peak observations. As one can see from the Tables, the equivalent  
widths measured from McDonald and Kitt Peak spectra agree very well. 
 
The listed uncertainties are only those associated with random noise in 
the continua and the width of the profile. Systematic uncertaintes due to  
continuum placement and other factors in the present dataset 
are not included. Typically, uncertainties in  
total $W_{\lambda}$ values caused by misplacements of continua are 5\%  
for weak lines and 2-3\% for strong ones. Considering these uncertainties,  
our total 
equivalent widths agree very well with previous measurements. Among our 228  
measurements, 100 have previous $W_{\lambda}$ values. Only 3 of them 
show statistically significant discrepancies: 
\ion{Ca}{2} toward HD206165 (248.1 vs. 321 m\AA); 
CN R(0) toward HD206165 (6.5  vs. $<$ 3 m\AA);  
and CN P(1) toward $\rho$ Oph C (3.3 vs. 9.2 m\AA). Three others,  
\ion{Ca}{1} toward HD208905, HD209339, and HD209481, indicate possible  
disagreement with absolute $W_{\lambda}$ differences of about 4.0 m\AA.  
 
Of the 3 discrepant cases, the difference in CN P(1) of $\rho$ Oph C is  
most likely due to errors in the previous measurement (Federman, unpublished).  
The measurement  
gave consistent equivalent widths for the stronger CN R(0) and R(1) lines  
but a higher $W_{\lambda}$ for P(1). Furthermore, the $W_{\lambda}$ value  
for P(1) is larger than the equivalent width for R(1), an unphysical  
situation. The other two differences involve HD206165. It is difficult to make  
a firm statement. The difference in \ion{Ca}{2} toward this star 
could be due to differences in recognizing the strong stellar line, on which 
the interstellar \ion{Ca}{2} feature is superposed. 
However, this can not explain 
the disagreement in CN.  It is possible that the 
differences may be caused by small-scale variation along 
the line of sight. The proper motion of HD206165 is about 3.4 mas yr$^{-1}$.  
The time span between our observations and the previous one  
(Chaffee \& Dunham 1979) is about 25 yrs, corresponding to an 85 AU  
separation at a distance of about 1000 pc. Evidence for pervasive small-scale  
structure on this scale has been found from ultraviolet, optical, and  
radio observations (e.g., Frail et al. 1994; Lauroesch \& Meyer 1999;  
Pan et al. 2001; Welty \& Fitzpatrick 2001). Furthermore, the previous  
measurement (Chaffee \& Dunham 1979) also reported a higher \ion{Ca}{1}  
$W_{\lambda}$ value, 12 m\AA, compared to ours of 7.8 m\AA.

\subsection{Component Structure} 
 
Our spectra clearly exhibit multiple velocity components in each 
line of sight, as seen in Figs. 1 to 15.  
Each interstellar line profile carries  
information about the number of velocity components, 
column density ($N$), the wavelength ($\lambda$) or velocity ($V$)
at line center, and line width ($b$-value, $\sim$ FWHM/1.665) for  
each component. We used the method of profile fitting 
to detemine column densities, 
line widths, and velocities of individual components 
contributing to the observed aborption  
line profile. The program FITS6P (Welty, Hobbs, \& York 1991) 
was used for this purpose. We determined the component structure by 
analyzing spectra of all the observed  
species for a given line of sight together. 
Therefore, our component structure of 
each species is also constrained by spectra of other 
species for the same line of sight. 
  
If only one component is present, the profile is assumed to be symmetric, and 
the determination of its $\lambda$, $b$-value, and $N$ is  
straightforward. The wavelength can be determined from the centroid of the  
line profile; $W_{\lambda}$ and line width can be measured by fitting a  
single Gaussian or Voigt profile.  Then the column 
density can be derived by using curves of growth.  
However, if a line profile consists of multiple components, the task of  
determining $\lambda$ and $W_{\lambda}$ of each component is considerably more  
difficult if the components are not completely resolved. The task could be  
even harder when the absorption features become saturated. Unfortunately, the  
latter is the case for most of the \ion{K}{1} and \ion{Ca}{2}  
measurements. Typically, there are more than 10 velocity components  
in a \ion{Ca}{2} profile and 7 in a \ion{K}{1} spectrum 
for each line of sight. Furthermore, the central components are usually quite 
saturated. Therefore, it is very difficult to extract the component structure  
from only \ion{K}{1} and \ion{Ca}{2} spectra. On the other hand, CN, CH, 
CH$^{+}$, and \ion{Ca}{1} spectra provide information on  $V_{LSR}$ for the  
strong \ion{K}{1} and \ion{Ca}{2} components because the distributions of 
different species in the same line of sight are related more or less. For  
instance, if there is a CN component at a certain $V_{LSR}$, we would expect  
a CH component at that $V_{LSR}$. The same is true for CH and \ion{K}{1}.  
Among our observed species, \ion{Ca}{2} is the most widely distributed. If  
any other species has a component at a specific $V_{LSR}$, presumably,  
\ion{Ca}{2} will have a corresponding component at that velocity. Thus,  
$V_{LSR}$ information for strong \ion{Ca}{2} components can be 
obtained by measuring spectra of other species for which the lines  
are not saturated. Usually, the central wavelengths of weak   
components on both shoulders of a \ion{Ca}{2} line profile can be determined 
relatively well by examining the line profile itself. Therefore, by analyzing  
spectra of all species together, component structure of each species is much  
better constrained than when it is determined from a spectrum for only one  
species. With a well-defined  
component structure, reliable column densities and line widths 
for each component can be extracted from the spectra.  
In all measurements, we adopted the minimum number of components needed to  
fit the observed line profiles adequately, given the SNR achieved in each case. 
In a few cases, velocity offsets from one species to another were recognized 
by a common difference for all components; they were treated in the fitting  
procedures.  Table 5 displays adopted component structures for all 
species; the LSR velocities given in column 1 represent an average from 
the suite of fits.
 
Different species may trace regions of a cloud where physical conditions  
are different, or sometimes even different clouds. 
For instance, CN traces dense gas. Some clouds are not dense enough to 
have detectable CN lines. As expected,  
among the six species, the CN profile is usually much less complicated than  
profiles of other species for a given sight line. This makes the CN spectrum a  
good starting point for the analysis.  
 
In our CN spectra, there are at most 5 velocity components for a given line  
of sight. They are usually well resolved and far from saturated. Thus, CN  
component structures can be determined adequately by profile fitting the CN  
spectrum itself in most cases. We started the analysis by examining the  
symmetry and width of the CN line profile. If the CN feature was 
symmetric, we first attempted one-component fitting. If the fitted  
profile was not very broad, we proceeded with one component 
fitting. If the fitted profile was too broad, i.e., its corresponding
Doppler parameter, $b$-value, is greater than our set limit of 1.6 km s$^{-1}$ 
(discussed below), we examined CH 
and \ion{K}{1} spectra for the sight line, and tried to find  
constraints from them. Usually we ended up with multiple  
(in most cases, two) CN components. An asymmetric line profile is a  
telltale sign of the presence of multiple components along a line of sight. 
If the CN profile is asymmetric, multiple component structure was adopted.  
The goodness of a fit was checked by examining the residuals. We ensured that  
residuals, after subtracting the fitted profiles, were indistinguishable from  
the noise in the continuum. The FITS6P program outputs $N$, $b$-value,  
velocity, and $W_{\lambda}$ of individual components. The second and 
third columns of Table 5 list the derived $N$ and $b$-value of each CN 
component.

It is rather difficult, if not impossible, to find the unique ``true'' component 
structure of a complicated profile from profile fitting because of spectrum 
noise, not high enough spectral resolution, and too many free parameters that we
have to fit. As stated above, we may be able to constrain the number of 
components by analyzing spectra of other species, 
but this is not always the case. Therefore, 
we set the following $b$-value limits for components of different species in our 
analysis. For CN, the limit was 1.6 km s$^{-1}$, 
which is two times mean $b$-values found by Gredel et al. (1991) 
and 2.5 times average $b$-values found by Crane et al. (1995). 
It was 2.0 km s$^{-1}$ for CH, \ion{Ca}{1}, and \ion{K}{1},
which is two times the $b$-values adopted by Knauth et al. (2001) and 3 times
median $b$-values obtained by Welty et al. (2003) and Welty and Hobbs (2001).
A limit for CH$^+$ and \ion{Ca}{2} of 3.5 km s$^{-1}$, 1.0 km s$^{-1}$ greater than
those adopted by Knauth et al. (2001), was chosen. We point out that
these limits played at most a minor role in our analysis because there were 
few instances requiring them.   
 
Once the CN component structure was obtained for a line of sight, we analyzed 
the CH profile for the sight line. First, we examined 
CH components that could be well defined by working on the CH spectrum itself.  
If all components could be adequately  
determined in this simple way, our next step was to check whether all CN  
components were present in the CH profile. If not, an iteration was  
performed to ensure the CH profile included all CN components. A 
$V_{LSR}$ difference of $\la$ 0.3 km s$^{-1}$ between CH and CN components was  
considered a match (i.e., the same component was detected in both  
species). In some cases, CH profiles were not so simple. Usually, only the  
central wavelengths (velocities) of weak components on the sides of a CH  
profile could be defined adequately. The strong CH components, whose  
corresponding CN are usually detected, were blended. Fortunately, we have  
$V_{LSR}$ (or central wavelength) for these strong CH components from CN, 
and we adopted velocities from CN for strong CH components as our initial 
input to FITS6P.  In some cases, we had to add additional CH components, 
which neither were seen in CN nor identified through examination of 
the CH profile, in order to fit the data adequately.  
The effects of $\Lambda$-doubling on the CH line was included in  
profile fitting. The inferred column densities and $b$-values of 
CH components are listed in the fourth and fifth columns of Table 5. 
   
The CH$^{+}$ and \ion{Ca}{1} structures were studied next because they are  
useful in interpreting \ion{K}{1} and \ion{Ca}{2} profiles. Usually,  CH$^{+}$  
and \ion{Ca}{1} lines are not saturated. With the aid of the CH component  
structure, CH$^{+}$ and \ion{Ca}{1} components were not too difficult to  
define. Although we fit line profiles of these two species so that they could  
share common velocity components with other species (analyzed to this point),  
none of the component structures for the two species was required to contain  
the structure for other species in its subset. 
The fitted $N$ and $b$-values for CH$^+$ and \ion{Ca}{1} 
components are in columns 6 to 9 of Table 5. 
 
\ion{K}{1} components were determined with the aid of the CH, CH$^{+}$, and  
\ion{Ca}{1} component structures. Because more components were present in  
\ion{K}{1} line profiles (as many as 10 in our spectra), their analysis is  
more difficult than for CH. In a few cases, we had to go through   
iterations from CN and CH to \ion{K}{1} profile fitting in order to get an  
adequate fit with reasonable widths, and to make the 
CH component structure a subset of the \ion{K}{1} structure for a given line  
of sight. HD207308 is one such case. The CN line profile for this line  
of sight looks ``symmetric''. We first attempted to  
fit it with one component. The one-component fitting yielded a fairly large  
line width with a {\it b}-value of $\sim$ 1.5 km s$^{-1}$. Since it was still  
slightly smaller than the {\it b}-value limit set for CN (1.6 km s$^{-1}$),  
we started the CH analysis with one CN component. However, we could not get 
consistent CN, CH, and \ion{K}{1} component structures. Only when the CN line  
profile was fit by two components did we get consistent results for all  
three species.  Moreover, the {\it b}-values of the two CN components  
dropped to more typical values, 1.0 and 0.9 km s$^{-1}$, respectively.  
Therefore, we feel more confident with this two-component structure. This  
analysis highlights how using spectra of several species together aids  
in obtaining reliable component structure for all species.  
The hyperfine splitting of \ion{K}{1} $\lambda$7699 was considered in 
the profile fitting. The derived \ion{K}{1} column density and $b$-value 
of each component are tabulated in the 10th and 11th columns of Table 5.  
 
The \ion{Ca}{2} line profile was the most difficult one to study, 
not only because more components were present, but also because the  
\ion{Ca}{2} components are more saturated. We see as many as 25 components  
in a \ion{Ca}{2} spectrum for a line of sight. In a few cases, the relative  
intensities of the \ion{Ca}{2} line profiles almost reached zero, as low as  
10$^{-3}$, though they never became black. The \ion{Ca}{2} spectra were fitted  
by using the same method as the one  
employed for the \ion{K}{1} measurements. However, in some cases,  
iterations were again performed to ensure consistent component  
structures for a given line of sight---i.e., velocity components seen in  
another species would appear in the \ion{Ca}{2} structure.  
Of course, \ion{Ca}{2} can have extra components not detected in any  
of the other observed species. The $N$ and $b$-values 
of \ion{Ca}{2} components are given in the last two columns of Table 5.   
 
As mentioned earlier, four spectra, \ion{K}{1} $\lambda$7699 toward 9 Cep,  
$\nu$ Cep, and $\lambda$ Cep, and \ion{Ca}{1} $\lambda$4226  
toward  $\lambda$ Cep, had much higher spectral resolution ($\sim$ 0.6 km  
s$^{-1}$) than others ($\sim$ 1.5 km s$^{-1}$). For consistency, we smoothed  
the higher resolution spectra before further analysis so that they had  
comparable resolution to the spectra of other species for these lines  
of sight.  
  
Because some \ion{K}{1} and \ion{Ca}{2} components were strong, even quite  
saturated, we could hardly extract useful information on how many nearby  
components are present by examining 
the \ion{K}{1} and \ion{Ca}{2} profiles alone.  
As stated above, we calculated the central wavelengths of these strong  
components using information from spectra of other species (such as CH$^{+}$,  
CH) in the same line of sight. Then we fit multiple Voigt profiles  
to the \ion{K}{1} and \ion{Ca}{2} line profiles by using the calculated  
central wavelength as initial input. However, there could be weak components  
(blended with strong ones) in line profiles not seen in other species. If 
this is the case, some closely blended weak components would be missed in our 
analyses. Since we did include some ``additional'' components in order to fit 
observed data adequately, we believe all components with reasonable strengths 
were found.   
 
One advantage of our study is that we observed six species for each line of  
sight with instruments having the same or similar resolutions, and reduced the  
data in the same fashion. By analyzing spectra of six species for a given 
line of sight together, we obtained consistent velocity component structures 
for all species. Therefore, we believe that the derived  
component structures, as well as the $N$ and $b$-value for each  
component, are more reliable than those deduced from a single species. 
 
There are factors that could introduce uncertainties into the  
derived column densities, $b$-values, and velocities. Uncertainties  
associated with noise in the spectra and placement of continua  
can be estimated whereas errors due to 
other factors, such as zero point offsets from one species to 
another in the profile fitting, may be rather difficult to quantify. 
For well defined intermediate strength components, the formal 1 $\sigma$ 
uncertainties in column densities are typicallly $\la$ 5\%, comparable 
to the uncertainties that would be inferred from the errors bars on  
equivalent widths associated with random noise in the spectra. The 
uncertainties in $N$ can be 10 to 20\% for strong but still  
well-defined components. As stated in $\S$3.1, misplacements of continua  
may cause 5\% uncertainties in equivalent widths for weak lines, and  
2---3\% for strong ones. For most components, the derived velocities and 
$b$-values have uncertainties of about 0.1 km s$^{-1}$, though they  
could be larger for broader and severely blended components.  
As noted above, component velocities derived from spectra of  
six observed species for a given line of sight agree within 0.3  
km s$^{-1}$, a conservative measure of the accuracy in velocity. In  
the process of extracting component structure, we usually let velocities 
and widths of components vary independently. However, in a few cases, 
the $b$-value of certain component(s) or the velocity of component(s) were 
set during the profile fitting.  In others, an offset from one species to  
another was recognized by a common difference for all components. 
These fitting restrictions and  
offset constraints could introduce errors in $N$ and $b$-value for  
some components. However, the $b$-values and velocities  
generally could not be changed by more than 0.2 km s$^{-1}$ and 0.3  
km s$^{-1}$, respectively, without noticably degrading the fit. 
There is also the possibility of unresolved structure.  In summary,  
uncertainties in our derived $N$ are $\la$ 20\% for most components, and 
somewhat larger for severely blended components. Typical uncertainties for  
$b$-values and velocities are 0.1 km s$^{-1}$. For significantly blended 
components, they are accurate to within 0.2 and 0.3 km s$^{-1}$, respectively. 
 
\subsection{Total Column Densities} 
 
Although we intend to interpret column density in term of velocity components  
rather than lines of sight, some studies based 
on lower resolution observations, 
such as {\it FUSE}, may need total column densities along lines of sight. The  
first column of each species in Table 6 lists  
our total column densities (hereafter FIT column density),  
based on summing the  
results of component structure along each line of sight. Comparison between  
these new total $N$ obtained from detailed profile fits and 
previous values from total equivalent widths provide information on how  
reliable those earlier values are, and provide a connection to the majority  
of existing data.  
 
Total column density along a line of sight can also be derived from the  
apparent optical depth method (Savage \& Sembach 1991). An apparent optical  
depth, $\tau_a$, is defined as a function of velocity,  
  
\begin{equation} 
 \tau_{a}(v) = ln\left[\frac{I_0{(v)}}{I(v)}\right], 
\end{equation}  
 
\noindent where $I(v)$ is the recorded intensity and $I_0(v)$ 
is the continuum level.  
If $\tau_{a}$ is not very large, the differential column density per unit  
velocity is expressed as  
 
\begin{equation} 
N_a(v) = 3.768\times 10^{14} 
\frac{\tau_a(v)}{f\lambda}~~~{\rm cm^{-2}}({\rm km\ s^{-1}})^{-1}, 
\end{equation}  
 
\noindent where $f$ is the transition's oscillator strength and $\lambda$ is  
the central wavelength in units of \AA.  
A direct integration of the differential column density, 
 
\begin{equation} 
 N_a =\int N_a(v)dv,  
\end{equation}  
   
\noindent yields an instrumentally smeared column 
density. In principle, this instrumentally smeared column 
density is a lower limit to the true column.  
Savage \& Sembach's (1991) simulations showed that the smeared  
column density is a good representation of the ``true value'' for weak and  
moderately strong absorption ($\tau \leq$ 0.5). For strong absorption lines,  
the instrumental resolution plays an important role: For a single component  
with a maximum $\tau$ =3.104 (relative intensity down to 0.045),  
their calculations showed that the smeared column density was smaller than  
the ``true value'' by 8.4\% when the instrumental resolution is  
1.6 times broader than line width ($b$-value). Their simulations also 
indicated that the smeared column density was even closer to the ``true value'' 
in multiple component cases for a given maximum $\tau$ and instrumental 
resolution.  
 
The apparent optical depth method makes no a priori assumptions regarding the  
velocity distribution of the absorbing gas, whereas profile fitting 
relies on component structures extracted from spectra of multiple  
species. As a check on our column densities  
and component structures, we obtained total column densities of each species  
using the apparent optical depth method for all sight lines  
(hereafter AOD column density). Results are listed in AOD columns of  
Table 6. FIT and AOD column densities for CN, CH, CH$^+$, and \ion{Ca}{1}
agree within 3\%, with FIT column densities generally slightly 
larger than their corresponding AOD column densities. 
For \ion{K}{1} and \ion{Ca}{2}, all FIT column densities are 
greater than AOD columns. The largest difference between 
FIT and AOD column densities are 22\% and 29\% for \ion{K}{1} and \ion{Ca}{2}, 
respectively. 
 
Although uncertainties in column densities for individual components can be 
20\%, uncertainties in total column densities 
for species for a line of sight are 
much smaller than that amount. Different component structure may cause  
column densities to change by 20\% for some individual components, but 
total column densities of all components can not be changed more than 5\% 
without noticably degrading the fit. Therefore, we conclude that uncertainties 
in total FIT column densities are about 5\%. 

For CN, CH, CH$^+$, and \ion{Ca}{1}, where absorption is not strong and our 
instrumental resolution is comparable to the inferred line widths, the lower  
limit column densities set by the apparent optical depth method 
should not be far from the ``true values'' according to 
Savage \& Sembach's (1991) 
simulations. Therefore, the fact that the FIT column densities agree with, but 
are generally slightly greater than, 
AOD densities indicates that our FIT column  
densities are good representations of their ``true 
values'' for these species, and the component structures on which FIT column  
densities are based are reliable. 

On the other hand, \ion{K}{1} and \ion{Ca}{2} absorption 
are usually much stronger, and their AOD column densities should be 
lower than their ``true values''. 
In Table 6, FIT column densities  
are typically greater than AOD columns by 15\% for \ion{K}{1}, and by 5\% for  
\ion{Ca}{2}, except for the four lines of sight toward $\rho$ Oph. 
While \ion{Ca}{2} lines are deeper than \ion{K}{1} lines, the AOD 
column densities for \ion{Ca}{2} are closer to their FIT 
values.  This reinforces our results that $b$-values 
for \ion{Ca}{2} components are greater than those of \ion{K}{1}.  
We note, however, that low SNR in the absorption cores 
of \ion{Ca}{2} profiles may increase  
AOD column densities. For instance, if 
the relative intensity in the core is 0.01 and SNR is 2, noise 
would increase the AOD column density by 15.1\% at some pixels 
and decrease it by 8.8\% at others.  The net effect is an 
increase of about 6.3\% in intergrated AOD column density.  Therefore, 
the differences between FIT and AOD column densities for 
\ion{Ca}{2} may be more like the difference seen in \ion{K}{1}.          
 
\section{SUMMARY} 
 
Interstellar absorption was observed along 29 lines of sight in three  
star-forming regions, $\rho$ Oph, Cep OB2, and Cep OB3. In this first paper,  
we presented high-resolution CN, CH, CH$^{+}$,  
\ion{Ca}{1}, \ion{K}{1}, and \ion{Ca}{2} spectra,  
describing the observations and data reduction. Total equivalent widths were   
compared with previous measurements; good agreement among determinations is  
the rule. The differences found for HD206165 may be the result of small-scale  
variation in the ISM. Our high-resolution spectra reveal complex component  
structure and closely blended components in most lines of sight.  
Reliable velocity component structure for all species 
was obtained by analyzing the spectra of each species for  
a given line of sight together. 
The column densities, $b$-values, and velocities of each component, 
determined by detailed profile fitting, are 
tabulated. Comparison of total column densities along lines 
of sight from the sums of component values from profile 
fitting and the apparent optical depth method (Savage \& Sembach 1991) 
highlighted the robustness of the fitting results. 
A more detailed analysis of  
these results and their implications will be presented in a companion  
paper (Pan et al. 2003). 
 
\acknowledgments 
 
It is our pleasure to thank the support staff of McDonald Observatory  
and KPNO, especially David Doss at McDonald Observatory and Daryl Willmarth 
at KPNO. This research made use of the Simbad database operated at CDS  
Strasbourg, France. K. P. acknowledges KPNO for providing board and lodging  
in Tucson during observing runs. We thank the 
anonymous referee for suggestions that have improved the paper. 
The work at the University of Toledo was  
supported by NASA grants NAG5--4957, NAG5--8961, and NAG5--10305 and grant 
GO--08693.03--A from the Space Telescope Science Institute. D.E.W. acknowledges 
support from NASA Long-Term Space Astrophysics grant NAG5-3228 to the 
University of Chicago. V.V.S. was funded through NASA Long Term Space  
Astrophyics grant NAG5-9213 to the University of Texas at El Paso.  
 
\appendix 
\section{Equivalent Width Measurements from Sandiford Data} 
In Table 7, we list $W_{\lambda}$ results from the Sandiford data set for  
those lines of sight that were not studied at the  high spectral resolution here.

\clearpage 
 
\begin{deluxetable}{lcccccccc} 
\tabletypesize{\scriptsize} 
\tablewidth{0pc} 
\tablecaption{Stellar Data for Program Stars} 
\tablehead{Star & Name & Sp. & $V$ & $l$ & $b$ & $E(B-V)$ & Distance & Spectra 
\tablenotemark{a} \\ 
      &    &   &(mag) & ($^\circ$)& ($^\circ$) &(mag) & (pc) & } 
\startdata 
\multicolumn{9}{c}{$\rho$ Oph}\\ 
HD147888  &$\rho$ Oph D &B4V&6.80& 353.65&+17.71&0.48 &150 &M01, M01, M01 \\ 
HD147932  &$\rho$ Oph C &B5V&7.27& 353.65&+17.71& 0.47 & 150 &M01, M01, M01  \\  
HD147933  &$\rho$ Oph A &B2V&5.02& 353.69&+17.69& 0.47 & 150& M01 \& K96, M01, M01 \& K96\\ 
HD147934  &$\rho$ Oph B &B1V& 5.92& 353.69&+17.69& 0.47 & 150 & M01, M01, M01 \\ 
\\ 
\multicolumn{9}{c}{Cep OB2}\\ 
HD203374A & & B0IVpe &6.69 & 100.51 & +8.62 & 0.60 & 800 & M00 \& K99, M00 \& K99, K00\\
\enddata 
\tablenotetext{a}{The first entry is for CN and Ca~{\tiny II}, the second for CH, 
CH$^+$, and Ca~{\tiny I}, the third for K~{\tiny I}. K99 = KPNO observations  
in 1999, M00 = McDonald spectra in 2000, and so on.} 
\end{deluxetable} 
 
\begin{deluxetable}{lcccccccc} 
\tabletypesize{\scriptsize} 
\tablewidth{0pc} 
\tablecaption{Total Equivalent Widths (m\AA) of CN Lines} 
\tablehead{Star&   & Ours & &  & & Others & & Ref. \\  
\cline{2-4} \cline{6-8}   
 (HD)     &R(0) & R(1)&  P(1) & & R(0) & R(1) & P(1) & } 
\startdata 
147888 &5.9$\pm$0.4 & 2.2$\pm$0.3 & 1.1$\pm$0.3 & &7.7$\pm$2.2 & $<$6.6&$<$6.6& 1 \\           
147932& 12.7$\pm$0.4& 5.1$\pm$0.4 & 3.3$\pm$0.4 & &15.9$\pm$1.9,7.4$\pm$2.1\tablenotemark{a} 
 & 6.4$\pm$1.4, $<$6.9\tablenotemark{a} & 9.2$\pm$2.5,$\ldots$ & 1, 2 \\ 
147933\tablenotemark{b} & 6.1$\pm$0.4 & 1.9$\pm$0.3 & 1.1$\pm$0.3 & & 6.1$\pm$0.5,5.9$\pm$0.3 &  
2.1$\pm$0.7,2.5$\pm$0.3 & 1.1$\pm$0.6, 1.2$\pm$0.7 & 3, 4\\  
      &7.0          &  2.4        & 2.0        &&               &             &             & \\ 
147934& 5.8$\pm$0.4 & 1.6$\pm$0.3 & 0.7$\pm$0.3 & & 6.8$\pm$0.5 & 2.1$\pm$0.7 & 1.3$\pm$0.7 & 3 \\  
203374A\tablenotemark{b} & 5.8$\pm$0.6  & 1.8$\pm$0.6 & 1.0$\pm$0.5 &  & 
$\ldots$   &$\ldots$    & $\ldots$   &      \\ 
          & 6.2$\pm$1.2  & $<$2.5 & $<$ 2.5  & &   &    &    &      \\ 
204827  &64.4$\pm$2.2 &30.7$\pm$2.0 & 18.2$\pm$1.8& &72.7$\pm$3.6 &31.1$\pm$2.8 &23.0$\pm$3.3 &5  \\
\enddata 
\tablenotetext{a}{$\rho$ Oph D instead of $\rho$ Oph C might have been observed.} 
\tablenotetext{b}{The first row are measurements from McDonald 
observations; the second row are results from KPNO spectra.} 
\end{deluxetable}  
 
\begin{deluxetable}{lccccccc} 
\tabletypesize{\scriptsize} 
\tablewidth{0pc} 
\tablecaption{Total Equivalent Widths (m\AA) of CH and CH$^+$ Lines} 
\tablehead{Star &    & CH  &    & &      & CH$^+$&  \\ \cline{2-4} \cline{6-8}    
(HD)       & Ours & Others & Ref. & & Ours & Others & Ref. } 
\startdata  
147888 &15.5$\pm$0.5 &18.0$\pm$0.9,17.6$\pm$3.1& 1,2 && 6.2$\pm$0.5 & 5.9$\pm$0.9 & 1\\  
       &             &14.0$\pm$0.6,16.3$\pm$0.5 &3,3& & & 7.2$\pm$0.8, 5.2$\pm$0.4&3,3\\ 
147932 &14.3$\pm$0.3 &14.1$\pm$2.8,14.8$\pm$0.9 & 1,2   && 6.0$\pm$0.6 & 8.7$\pm$1.6  & 2 \\          
147933\tablenotemark{a} &16.9$\pm$0.4 &16.9, 16.7$\pm$0.5 & 4,5 && 12.8$\pm$0.4 & 12.8$\pm$0.6, 
15.8,11.1$\pm$0.7,& 3,4,6\\  
       &             &       &      && 12.6 &    & \\ 
147934 &16.2$\pm$0.5 &$\ldots$&     && 11.5$\pm$0.6  &$\ldots$ &  \\  
203374A\tablenotemark{a} &17.3$\pm$1.0 &16.5$\pm$0.5 & 7 && 5.5$\pm$0.8 & 5.6$\pm$1.2 & 7\\ 
          &14.8$\pm$1.7 &   &  && 4.3$\pm$1.5 &  & \\
\enddata 
\tablenotetext{a}{The first row are measurements from McDonald observations; 
the second row are results from KPNO spectra.} 
\end{deluxetable}

\begin{deluxetable}{lccccccccccc} 
\tabletypesize{\scriptsize} 
\hoffset= -0.35 in 
\tablewidth{0pc} 
\tablecaption{Total Equivalent Widths (m\AA) of \ion{Ca}{1}, \ion{Ca}{2}, 
 and \ion{K}{1} lines} 
\tablehead{Star &   &Ca~{\tiny I}&  &&   &Ca~{\tiny II}&  &&  &K~{\tiny I}& \\  
   \cline{2-4} \cline{6-8} \cline{10-12}   
(HD) & Ours & Others & Ref. && Ours & Others & Ref.& & Ours & Others & Ref.} 
\startdata 
147888 & 5.5$\pm$0.5 &  6.4$\pm$0.9,  5.8$\pm$0.4 & 1,2 & 
       &63.6$\pm$3.0 & 60.7$\pm$2.4               & 1 & 
       &87.9$\pm$3.0 & 92.0$\pm$2.0, 87.3$\pm$0.6 & 1,2 \\ 
       &             & $\ldots$                   &   &  
       &             & 54.8$\pm$1.6, 54.7$\pm$2.9 & 2,2 & 
       &             & 85.6$\pm$1.2               & 2 \\   
 
147932 & 4.7$\pm$0.5 &  3.1$\pm$0.7               & 1 & 
       &61.1$\pm$2.0 & 59.7$\pm$3.6               & 1 & 
       &77.8$\pm$3.0 & $\ldots$                   &   \\ 
 
147933\tablenotemark{a} & 
         4.7$\pm$0.3 &  5.0$\pm$2.0,  4.5$\pm$0.6 & 3,4 & 
       &62.4$\pm$2.0 & $\ldots$                   &   & 
       &85.7$\pm$2.0 & 87,91$\pm$4                & 5,6 \\      
       & 5.7$\pm$0.6 &  4.8$\pm$0.4               & 8 & 
       &57.9$\pm$1.1 & $\ldots$                   &   &  
       &87.3$\pm$1.2 & 86.0$\pm$1.6, 85.4$\pm$0.8 & 2,10 \\         
 
147934 & 4.5$\pm$0.4 & $\ldots$                   &   & 
       &63.7$\pm$2.5 & $\ldots$                   &   & 
       &85.7$\pm$3.0 & 85.0                       & 5 \\ 
 
203374A\tablenotemark{a} & 
         8.1$\pm$1.0 & 10.0$\pm$1.5               & 7 & 
      &266.9$\pm$4.0 & $\ldots$                   &   & 
      &$\ldots$      & $\ldots$                   &   \\ 
      &  7.9$\pm$1.8 & $\ldots$                   &   & 
      &266.5$\pm$7.0 & $\ldots$                   &   & 
      &148.9$\pm$4.0 & $\ldots$                   &   \\
\enddata 
 
\tablerefs{ 
 1) Federman unpublished;  
 2) Welty unpublished; 
 3) Snow et al. 1986;  
 4) Lambert \& Danks 1986; 
 5) Lauroesch \& Meyer 1999;  
 6) Chaffee \& White 1982; 
 7) Sandiford data;  
 8) Welty et al. 2003;  
 9) Chaffee \& Dunham 1979;  
10) Welty \& Hobbs 2001;  
11) Allen 1994; 
12) Hobbs 1984;  
13) Munari \& Zwitter 1997.} 
 
\tablenotetext{a}{The first row are measurements from McDonald observations; 
the second row are results from KPNO spectra.} 
\tablenotetext{b}{Components at $V_{LSR}$ = -35 to -70 km s$^{-1}$ 
might not be included in the measurement.} 
 
\end{deluxetable}

\begin{deluxetable}{rccccccccccccccccc} 
\tabletypesize{\scriptsize} 
\tablewidth{0pc} 
\tablecaption{Component Measurements of Interstellar Lines} 
\tablehead{ &\multicolumn{2}{c}{CN} && \multicolumn{2}{c}{CH} &&\multicolumn{2}{c}{CH$^+$}&& 
\multicolumn{2}{c}{Ca~{\tiny I}} &&\multicolumn{2}{c}{K~{\tiny I}} &&\multicolumn{2}{c}{Ca~{\tiny II}}\\  
   \cline{2-3} \cline{5-6} \cline{8-9} \cline{11-12} \cline{14-15} \cline{17-18} 
$V_{LSR}$\tablenotemark{a}& $N$\tablenotemark{b} & $b$\tablenotemark{a} && $N$\tablenotemark{b} & $b$\tablenotemark{a}  
&& $N$\tablenotemark{b} & $b$\tablenotemark{a} && $N$\tablenotemark{b} & $b$\tablenotemark{a} && $N$\tablenotemark{b}  
& $b$\tablenotemark{a} && $N$\tablenotemark{b} & $b$\tablenotemark{a} \\ 
 &(10$^{12}$) &  & &(10$^{12}$) &  &&(10$^{12}$) &  &&(10$^{9}$) &  & 
         &(10$^{11}$) &  & &(10$^{11}$) &  } 
\startdata 
    \multicolumn{18}{c}{HD147888}\\ 
-20.6&$\ldots$  &   && $\ldots$ & &&$\ldots$ &   && $\ldots$&   &&$\ldots$& && 0.2 & 1.9\\      
-2.8&$\ldots$  &   && $\ldots$ & &&$\ldots$ &   && $\ldots$&   &&0.1 & 0.9&& 0.4 & 1.5\\     
0.2&$\ldots$  &   && 2.3 & 1.0 &&1.0 &1.7&& 2.4 &1.0&& 0.9 & 1.0&& 2.4 & 1.0\\     
2.0&2.1 &0.8&&19.5 &0.9&&6.5 & 2.0&&17.8 & 1.2&& 9.4 & 1.0&&14.2 &1.1\\ 
3.6&$\ldots$  &   && $\ldots$ & &&$\ldots$ &   && $\ldots$&   &&0.6 & 0.7&& 2.2 & 1.0\\  
\\ 
 \multicolumn{18}{c}{HD147932}\\
\enddata 
\tablenotetext{a}{in the unit of km s$^{-1}$.} 
\tablenotetext{b}{in the unit of cm$^{-2}$.} 
\tablenotetext{c}{No available CH$^+$ and Ca~{\tiny I} data for this line of sight.} 
\tablenotetext{d}{Ca~{\tiny II} components at $V_{LSR} < $ 40 km s$^{-1}$ 
are not included.} 
\tablenotetext{e}{No available CH and CH$^+$ data for this line of sight.} 
 
\end{deluxetable} 
 
\clearpage 
 
\begin{deluxetable}{lccccccccccccccccc} 
\tabletypesize{\scriptsize} 
\tablewidth{0pc} 
\tablecaption{Total Column Densities} 
\tablehead{& \multicolumn{2}{c}{CN} & & \multicolumn{2}{c}{CH} && \multicolumn{2}{c}{CH$^+$} && 
\multicolumn{2}{c}{Ca~{\tiny I}} && \multicolumn{2}{c}{Ca~{\tiny II}} &&  
\multicolumn{2}{c}{K~{\tiny I}} \\ 
Star & \multicolumn{2}{c}{($\times$10$^{12}$ cm $^{-2}$)} &&\multicolumn{2}{c}{($\times$10$^{12}$ cm $^{-2}$)}&& 
\multicolumn{2}{c}{($\times$10$^{12}$ cm $^{-2}$)} && \multicolumn{2}{c}{($\times$10$^{9}$ cm $^{-2}$)}&&  
\multicolumn{2}{c}{($\times$10$^{11}$ cm $^{-2}$)} & &\multicolumn{2}{c}{($\times$10$^{11}$ cm $^{-2}$)}\\  
 \cline{2-3} \cline{5-6} \cline{8-9} \cline{11-12} \cline{14-15} \cline{17-18}  
 & FIT & AOD && FIT & AOD && FIT & AOD && FIT & AOD && FIT & AOD && FIT & AOD } 
\startdata 
HD147888  & 2.1 & 2.1 && 21.8 & 21.2 && 7.5 & 7.2 && 20.2 &19.8  &&19.4 &14.9 &&11.0 &8.7 \\ 
HD147932  & 6.0 & 5.9 && 19.2 & 19.2 && 7.0 & 6.7 && 16.2 &16.5  && 19.2&15.5 && 8.0 & 6.7  \\  
HD147933  & 2.1 & 2.1 && 23.7 & 23.6 &&15.9 & 15.7 && 17.4&17.6 && 17.5 & 14.1&&10.3 &8.4  \\ 
HD147934  & 2.0 & 1.9 && 23.3 & 23.2 &&14.3 & 14.1&& 16.9& 17.0& & 19.6 & 13.9&&10.9 & 8.6 \\ 
HD203374A & 2.2 &2.1 && 22.2 &22.0 && 7.1 & 7.0 && 30.4 & 29.8 && 66.7 & 64.5 && 13.9& 11.8 \\ 
HD204827  &37.4 &37.2 && 87.9 & 86.4 && 36.9& 37.3 && 25.0 & 23.8 && 61.9 & 60.8 && 42.0 & 36.4 \\ 
HD206165  & 2.4 & 2.4 && 20.5 & 20.7 &&15.9 & 15.5 && 29.5 & 28.9 && 62.0 & 58.8 && 16.5 & 14.3 \\ 
HD206183  & 1.0 & 1.0 && 18.2 & 18.1 &&15.1 & 14.8 && 30.1 & 29.0 && 52.8 & 51.0 && 15.2 & 14.1\\
\enddata 
\end{deluxetable} 
 
\clearpage 
 
\begin{deluxetable}{lcccccc} 
\tablewidth{0pc} 
\tablecaption{Results from Sandiford Data} 
\tablehead{Star & $W_{\lambda}({\rm CH})$ & $V_{LSR}$ & 
$W_{\lambda}({\rm CH^+})$ & $V_{LSR}$ & 
$W_{\lambda}({\rm Ca~{\small I}})$ & $V_{LSR}$ \\  
 & (m\AA) & (km s$^{-1}$) & (m\AA) & (km s$^{-1}$)  
& (m\AA) & (km s$^{-1}$)} 
\startdata 
HD239649 & $\ldots$ & $\ldots$ & $10.4 \pm 0.8$ & $-9.4$ &  
$5.5 \pm 0.8$ & $-9.2$ \\ 
 & $12.3 \pm 1.1$ & $-0.6$ &  
$13.7 \pm 0.8$ & $+4.6$ & $4.1 \pm 0.8$ & $+1.3$ \\ 
HD239676 & $16.5 \pm 1.0$ & $-2.8$ &  
$13.0 \pm 0.9$ & $-1.5$ & $5.1 \pm 1.2$ & $-3.7$ \\ 
 & $20.7 \pm 1.0$ & $+3.4$ &  
$7.0 \pm 0.9$ & $+5.7$ & $6.1 \pm 1.2$ & $+5.0$ \\ 
HD239681 & $14.4 \pm 1.5$ & $-0.1$ &  
$22.2 \pm 1.5$ & $-3.4$ & $4.0 \pm 1.7$ & $-0.6$ \\ 
HD205794 & $22.3 \pm 0.8$ & $-8.2$ &  
$35.7 \pm 1.2$ & $-8.9$ & $2.0 \pm 0.6$ & $-5.4$ \\ 
 & $10.7 \pm 0.9$ & $+1.3$ &  
$7.5 \pm 1.1$ & $+3.0$ & $5.5 \pm 0.6$ & $+3.0$ \\ 
HD239710 & $22.9 \pm 1.1$ & $+1.9$ &  
$8.0 \pm 1.1$ & $+0.9$ & $4.6 \pm 1.4$ & $+1.8$ \\
\enddata 
\tablenotetext{a}{Line contaminated by cosmic ray hit.} 
\tablenotetext{b}{Possibly a stellar feature.} 
\end{deluxetable} 
 
\clearpage 

\begin{figure} 
\plotone{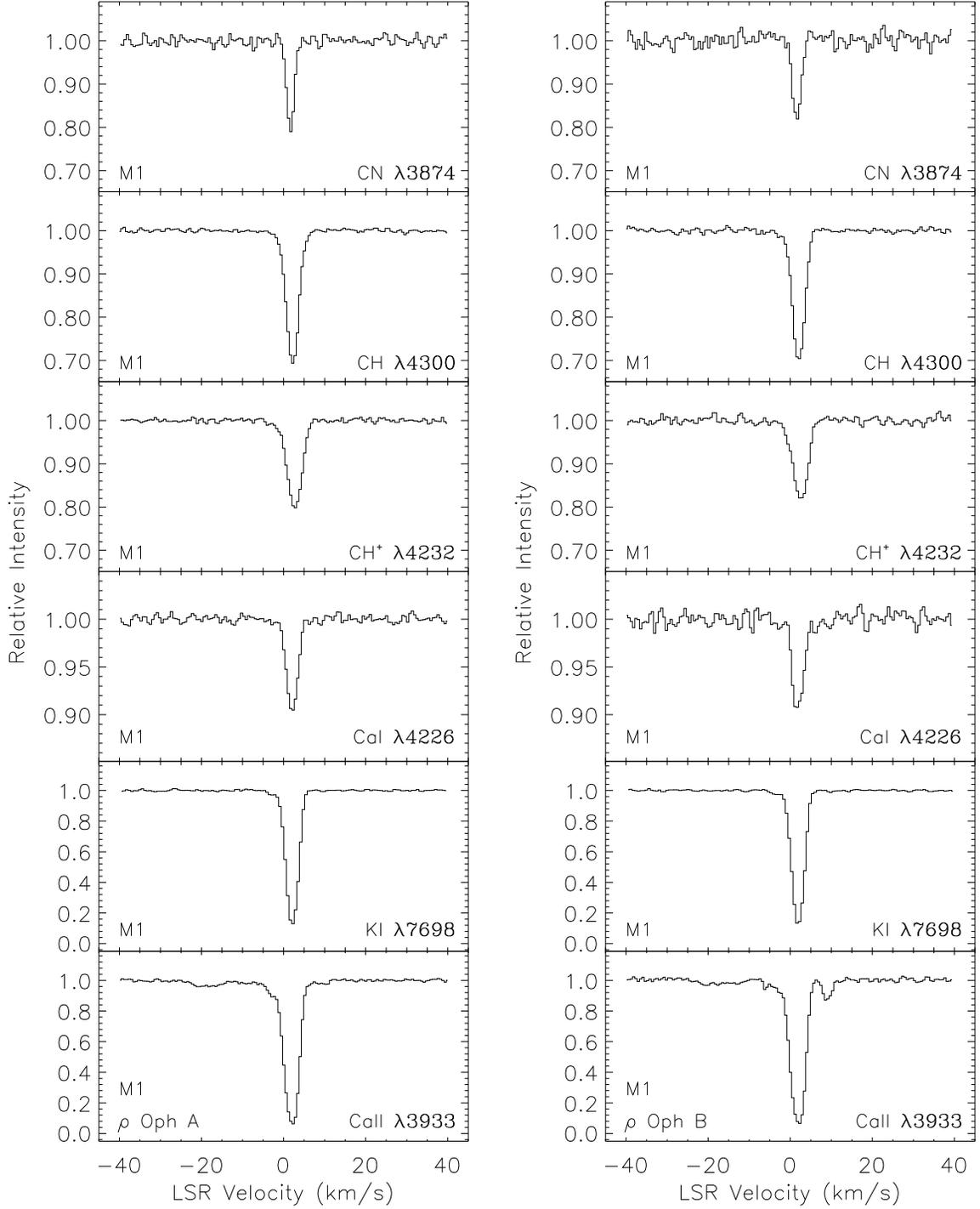} 
\caption{Interstellar CN, CH, CH$^+$, \ion{Ca}{1}, \ion{K}{1}, and 
  \ion{Ca}{2} absorption profiles toward $\rho$ Oph A 
  and $\rho$ Oph B. The transition is given in the lower-right corner  
  of each panel. The source of the spectra is denoted M1, M2, and K. M1  
  represents observations from McDonald Observatory during 2000 
and 2001, whereas M2 denotes observations in 1995 and 1996 from McDonald  
Observatory with a higher resolution. All KPNO spectra are denoted K.  
Note that the vertical scales differ from panel to panel.} 
\end{figure} 
 
\clearpage 
 
\begin{figure} 
\plotone{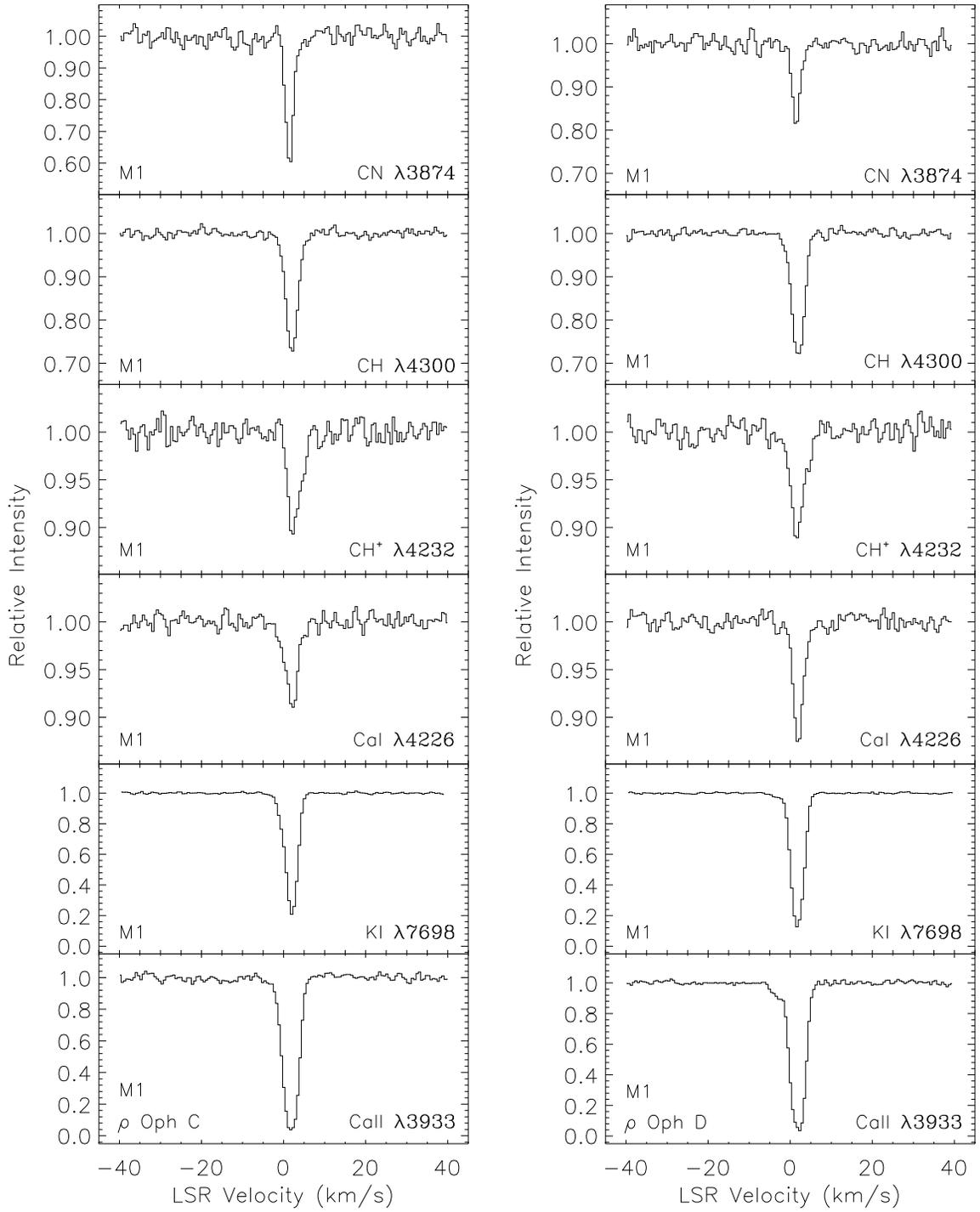} 
\caption{Interstellar CN, CH, CH$^+$, \ion{Ca}{1}, \ion{K}{1}, and 
  \ion{Ca}{2} absorption profiles toward $\rho$ Oph C 
  and $\rho$ Oph D (as for Figure 1).} 
\end{figure} 
 
\clearpage 
 
\begin{figure} 
\plotone{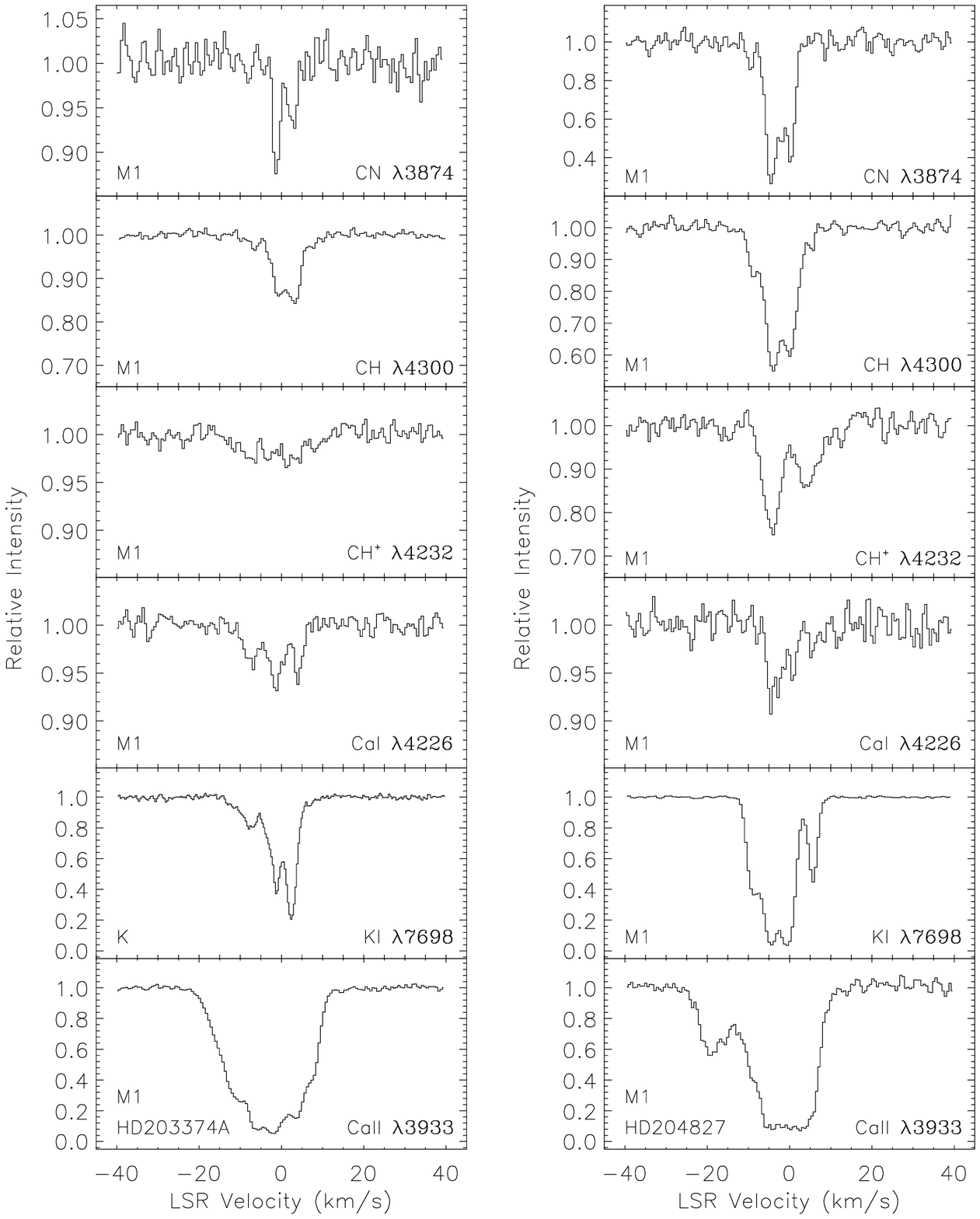} 
\caption{Interstellar  CN, CH, CH$^+$, \ion{Ca}{1}, \ion{K}{1}, and 
  \ion{Ca}{2} absorption profiles toward HD203374A 
  and HD204827 (as for Figure 1).} 
\end{figure} 
 
\clearpage 
 
\begin{figure} 
\plotone{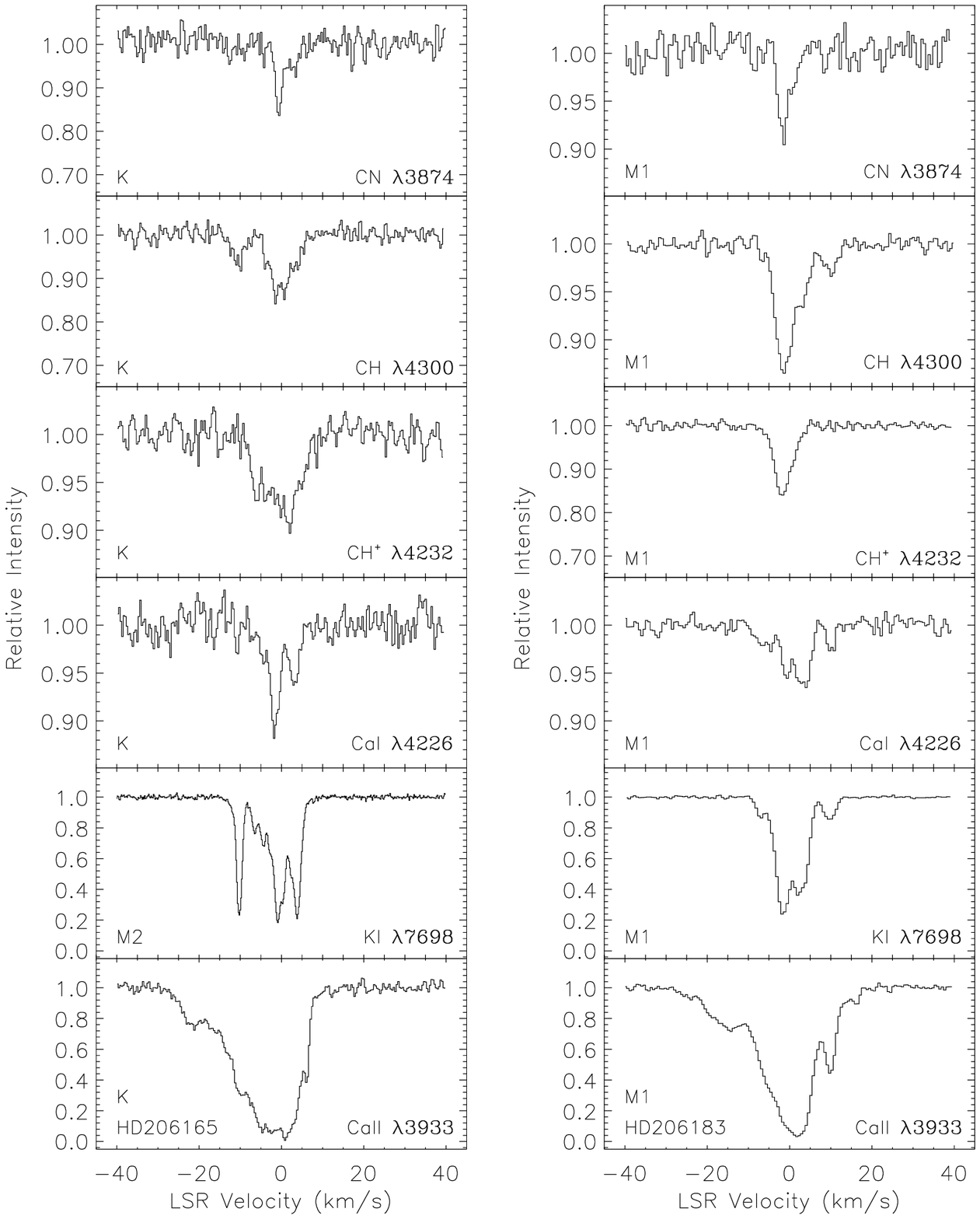} 
\caption{Interstellar  CN, CH, CH$^+$, \ion{Ca}{1}, \ion{K}{1}, and 
  \ion{Ca}{2} absorption profiles toward HD206165 
  and HD206183 (as for Figure 1).} 
\end{figure} 
 
\clearpage 
 
\begin{figure} 
\plotone{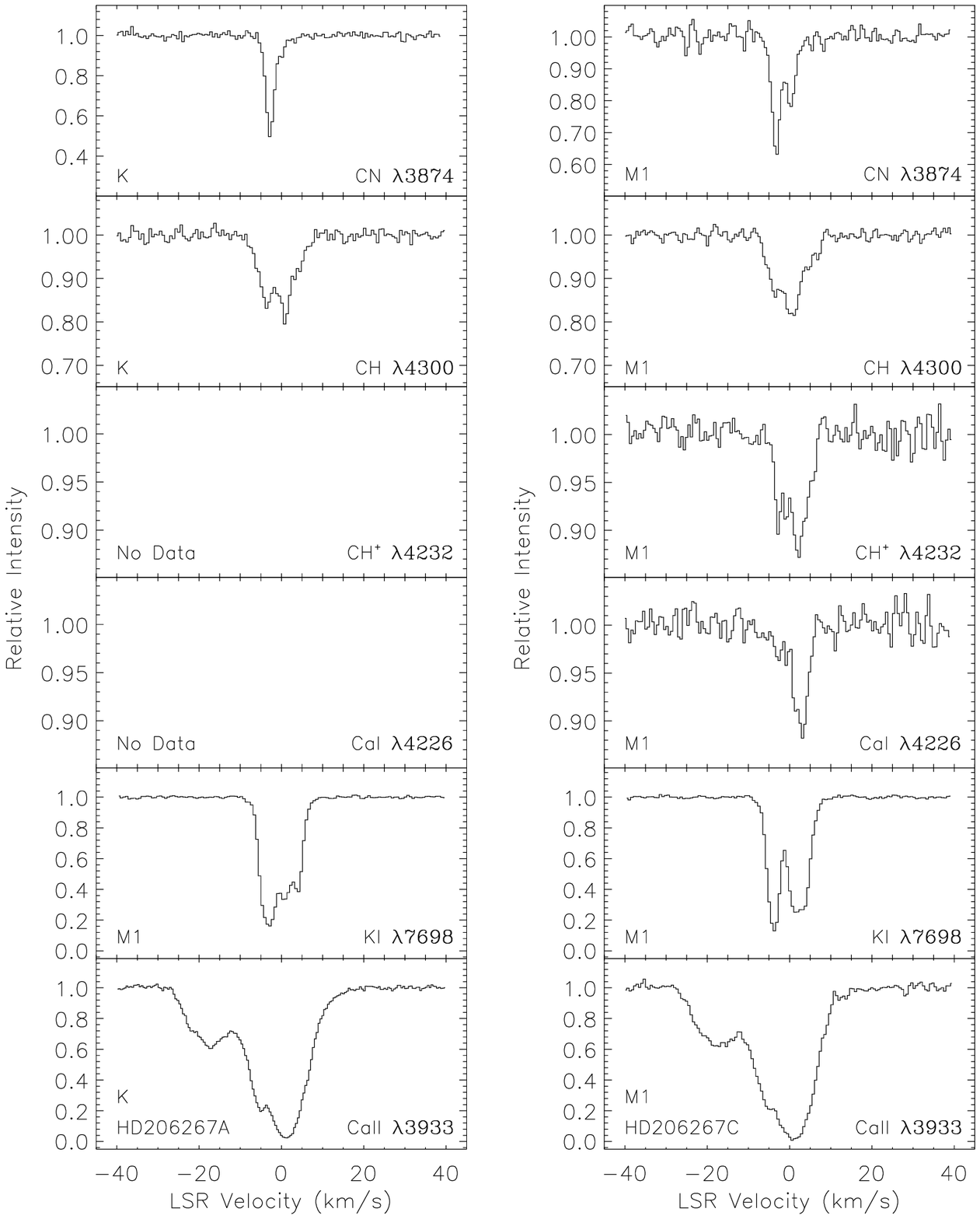} 
\caption{Interstellar  CN, CH, CH$^+$, \ion{Ca}{1}, \ion{K}{1}, and 
  \ion{Ca}{2} absorption profiles toward HD206267A 
  and HD206267C (as for Figure 1).} 
\end{figure} 
 
\clearpage 
 
\begin{figure} 
\plotone{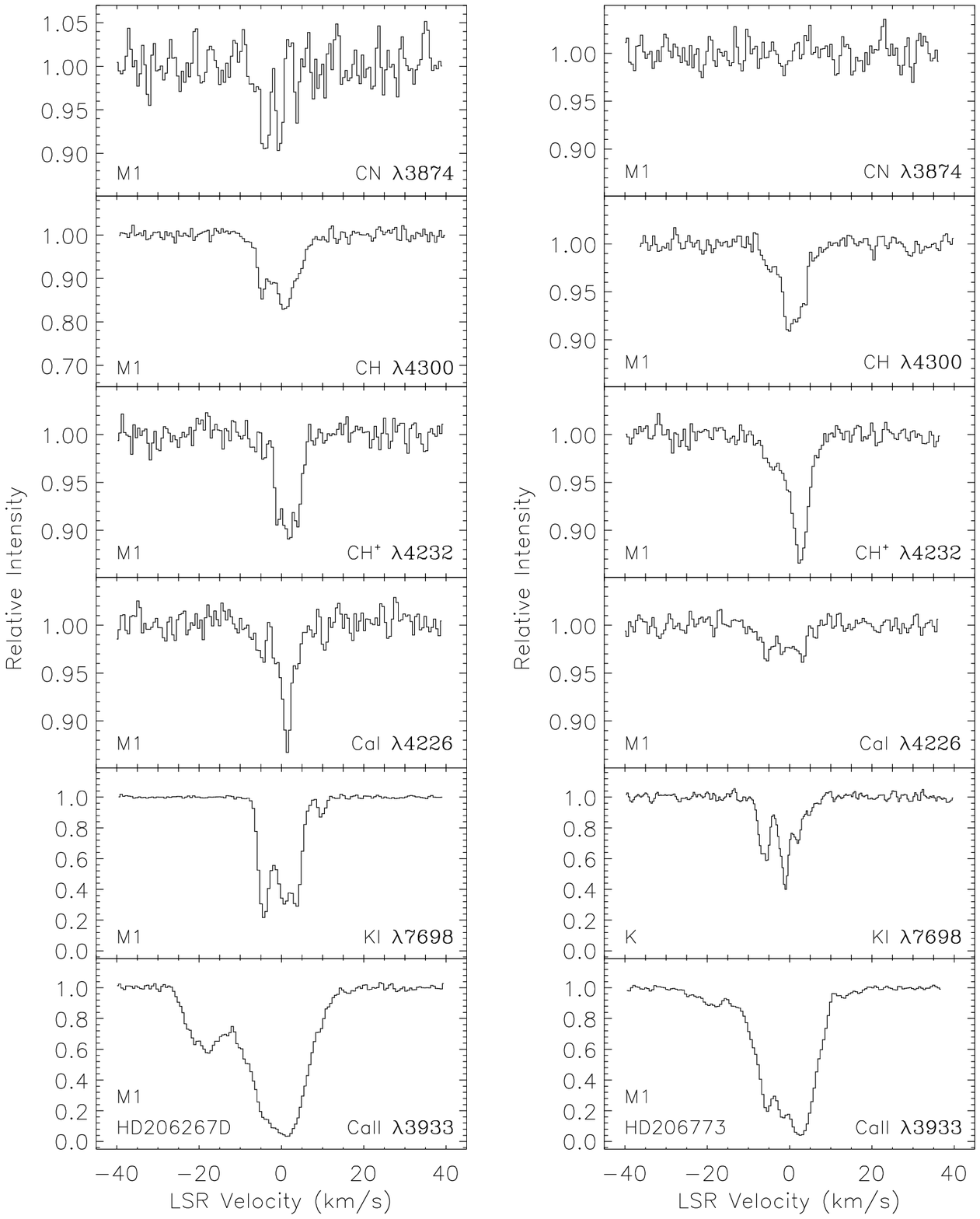} 
\caption{Interstellar  CN, CH, CH$^+$, \ion{Ca}{1}, \ion{K}{1}, and 
  \ion{Ca}{2} absorption profiles toward HD206267D 
  and HD206773 (as for Figure 1).} 
\end{figure} 
 
\clearpage 
 
\begin{figure} 
\plotone{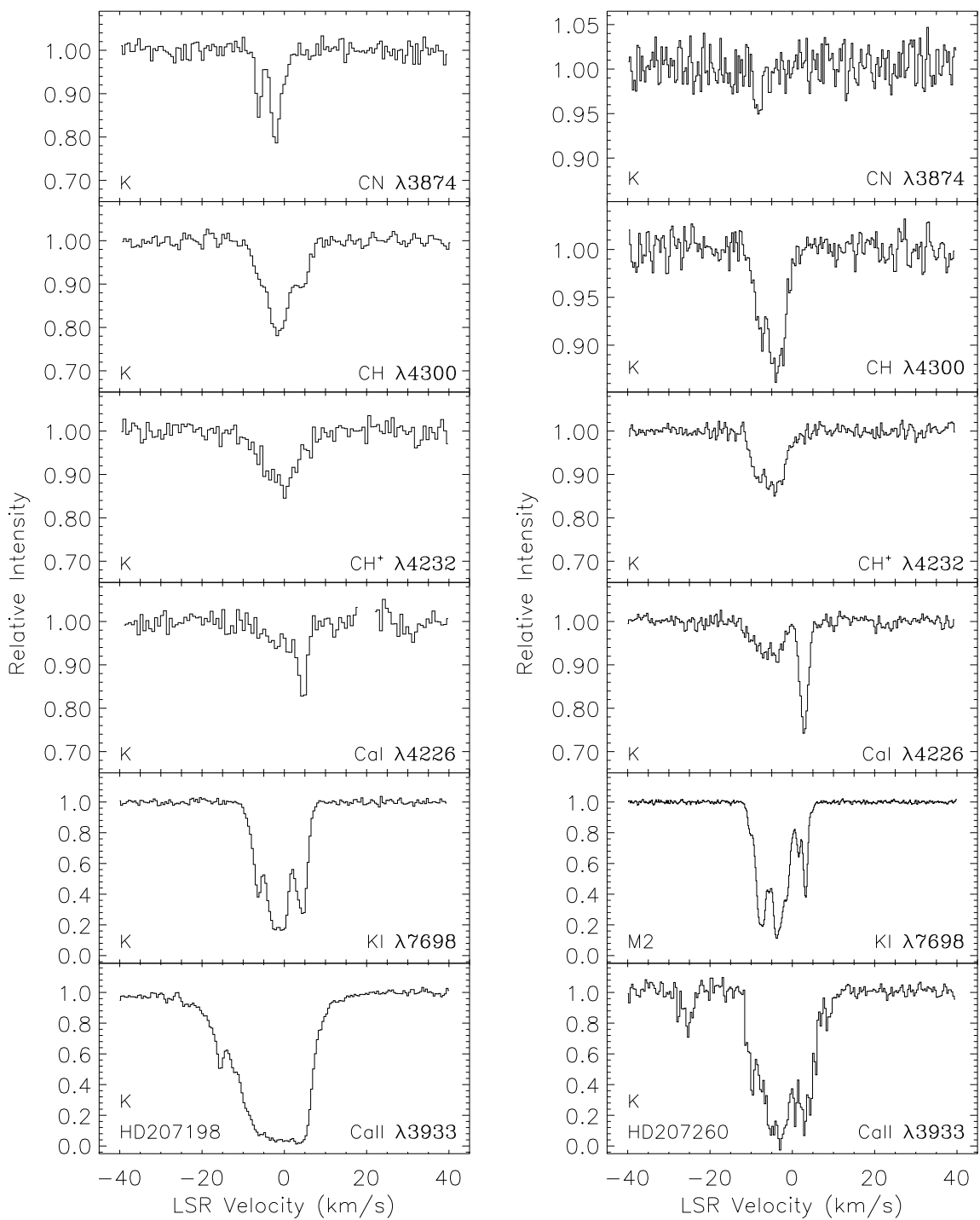} 
\caption{Interstellar  CN, CH, CH$^+$, \ion{Ca}{1}, \ion{K}{1}, and 
  \ion{Ca}{2} absorption profiles toward HD207198 
  and HD207260 (as for Figure 1). Blank region for \ion{Ca}{1} toward 
  HD207198 contains bad pixels.} 
\end{figure} 
 
\clearpage 
 
\begin{figure} 
\plotone{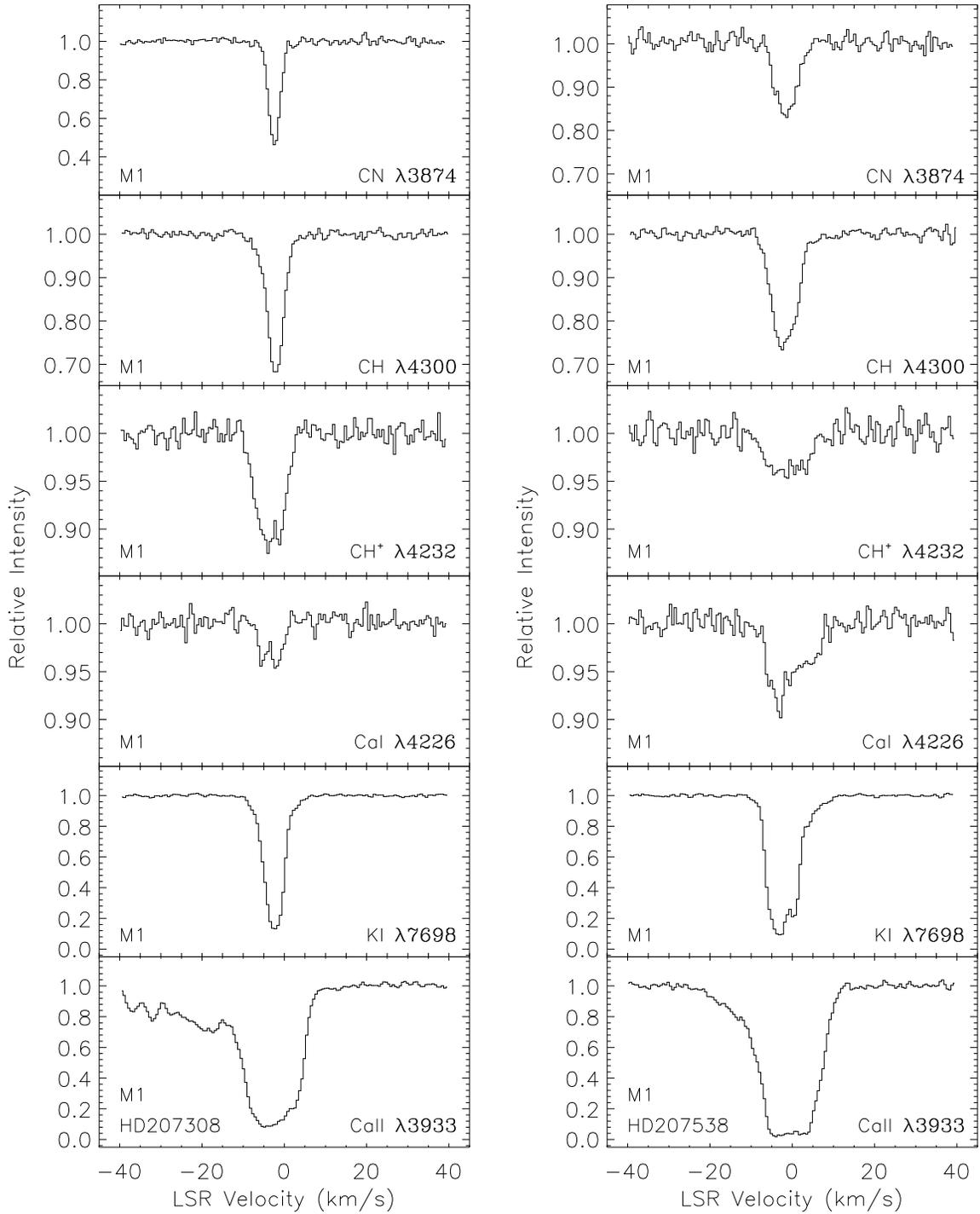} 
\caption{Interstellar  CN, CH, CH$^+$, \ion{Ca}{1}, \ion{K}{1}, and 
  \ion{Ca}{2} absorption profiles toward HD207308 
  and HD207538 (as for Figure 1).} 
\end{figure} 
 
\clearpage 
 
\begin{figure} 
\plotone{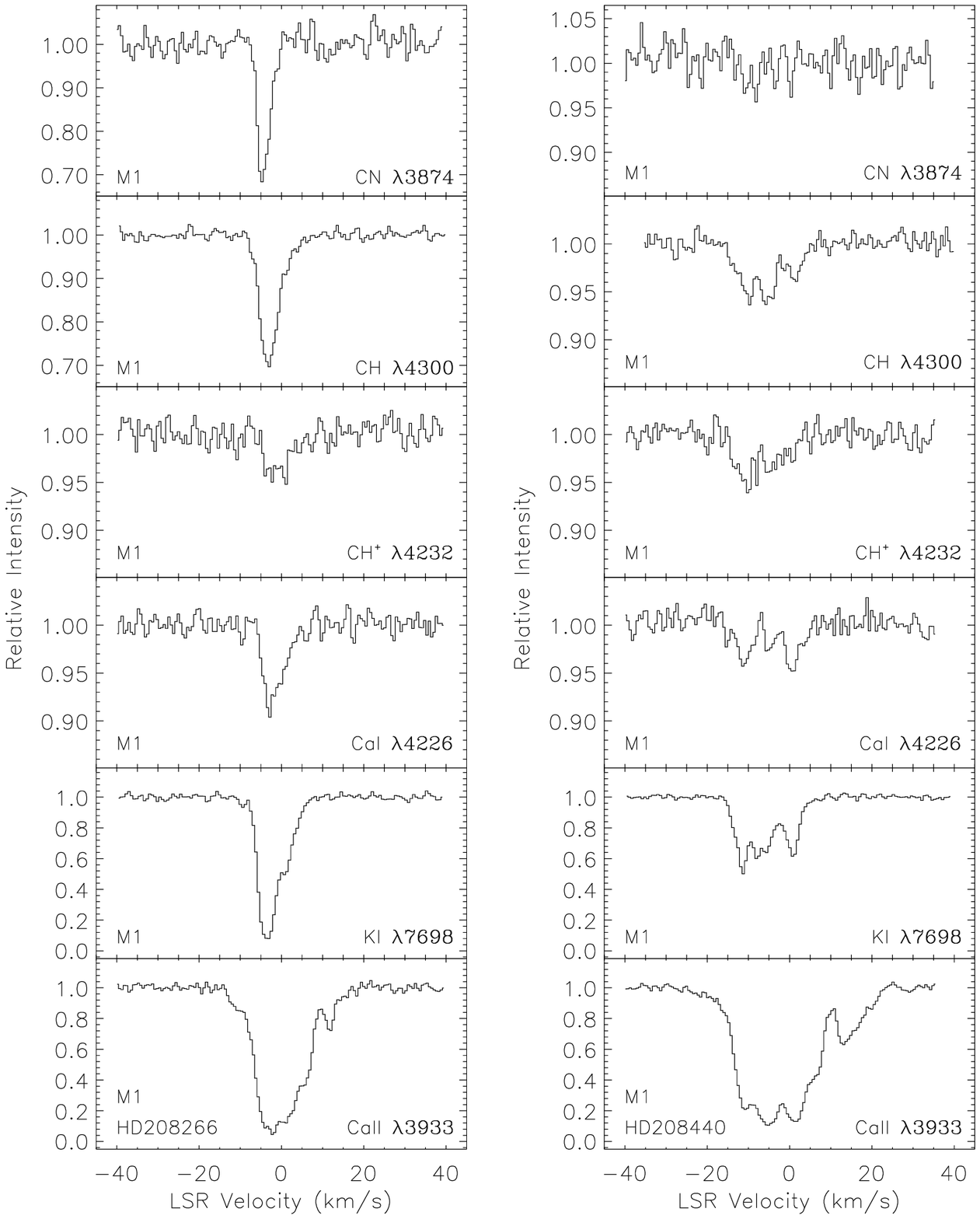} 
\caption{Interstellar  CN, CH, CH$^+$, \ion{Ca}{1}, \ion{K}{1}, and 
  \ion{Ca}{2} absorption profiles toward HD208266 
  and HD208440 (as for Figure 1).} 
\end{figure} 
 
\clearpage 
 
\begin{figure} 
\plotone{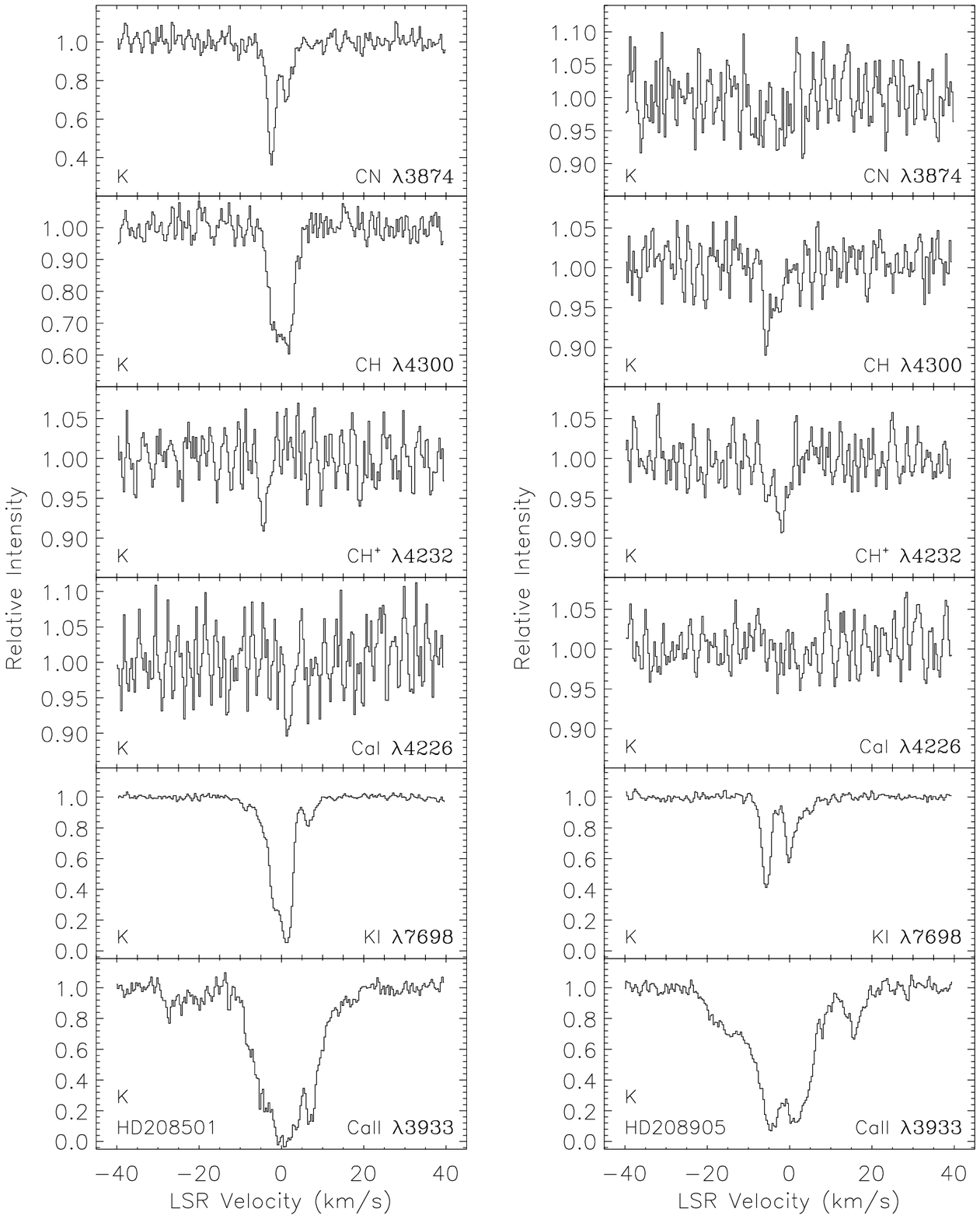} 
\caption{Interstellar  CN, CH, CH$^+$, \ion{Ca}{1}, \ion{K}{1}, and 
  \ion{Ca}{2} absorption profiles toward HD208501 
  and HD208905 (as for Figure 1).} 
\end{figure} 
 
\clearpage 
 
\begin{figure} 
\plotone{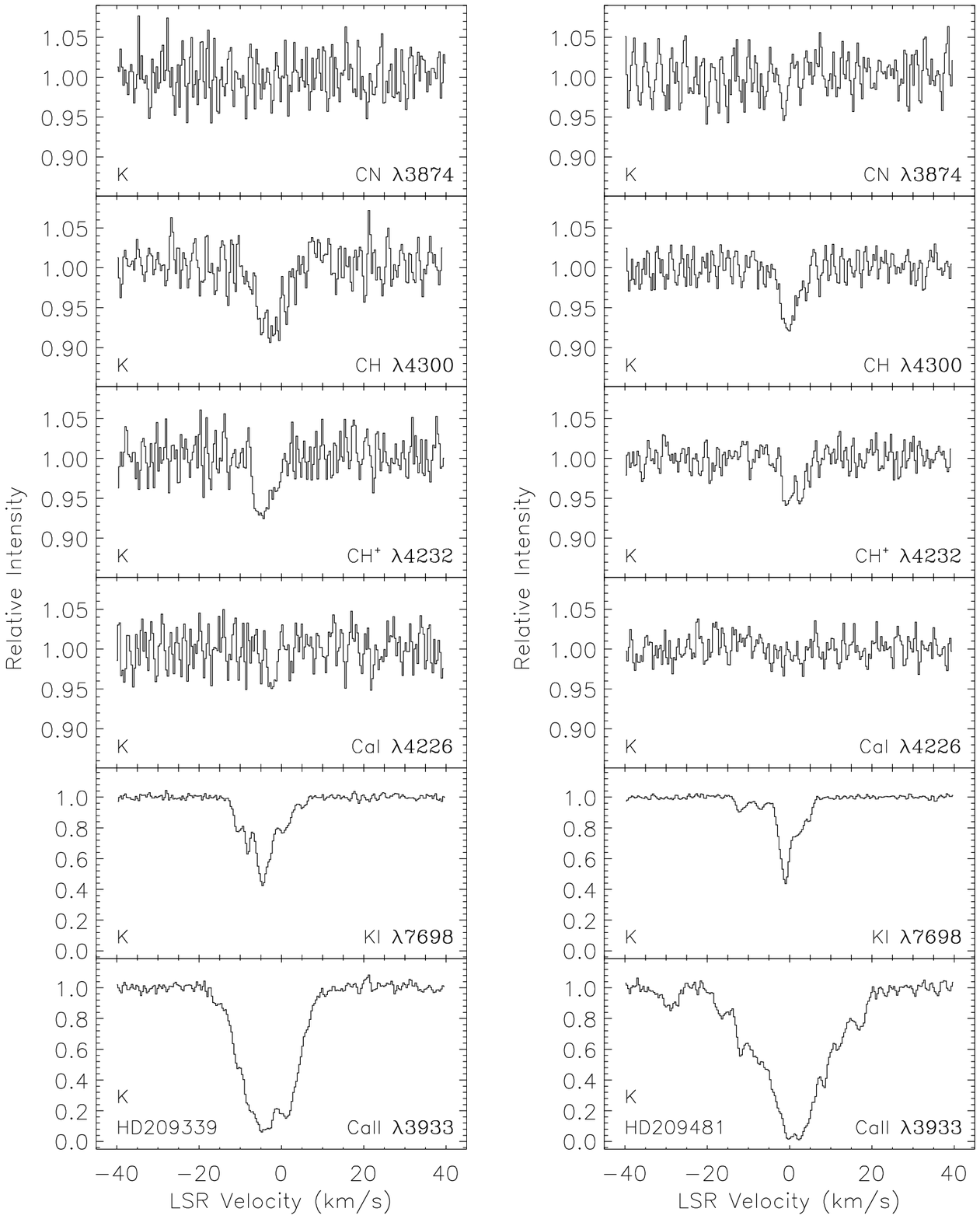} 
\caption{Interstellar  CN, CH, CH$^+$, \ion{Ca}{1}, \ion{K}{1}, and 
  \ion{Ca}{2} absorption profiles toward HD209339 
  and HD209481 (as for Figure 1).} 
\end{figure} 
 
\clearpage 
 
\begin{figure} 
\plotone{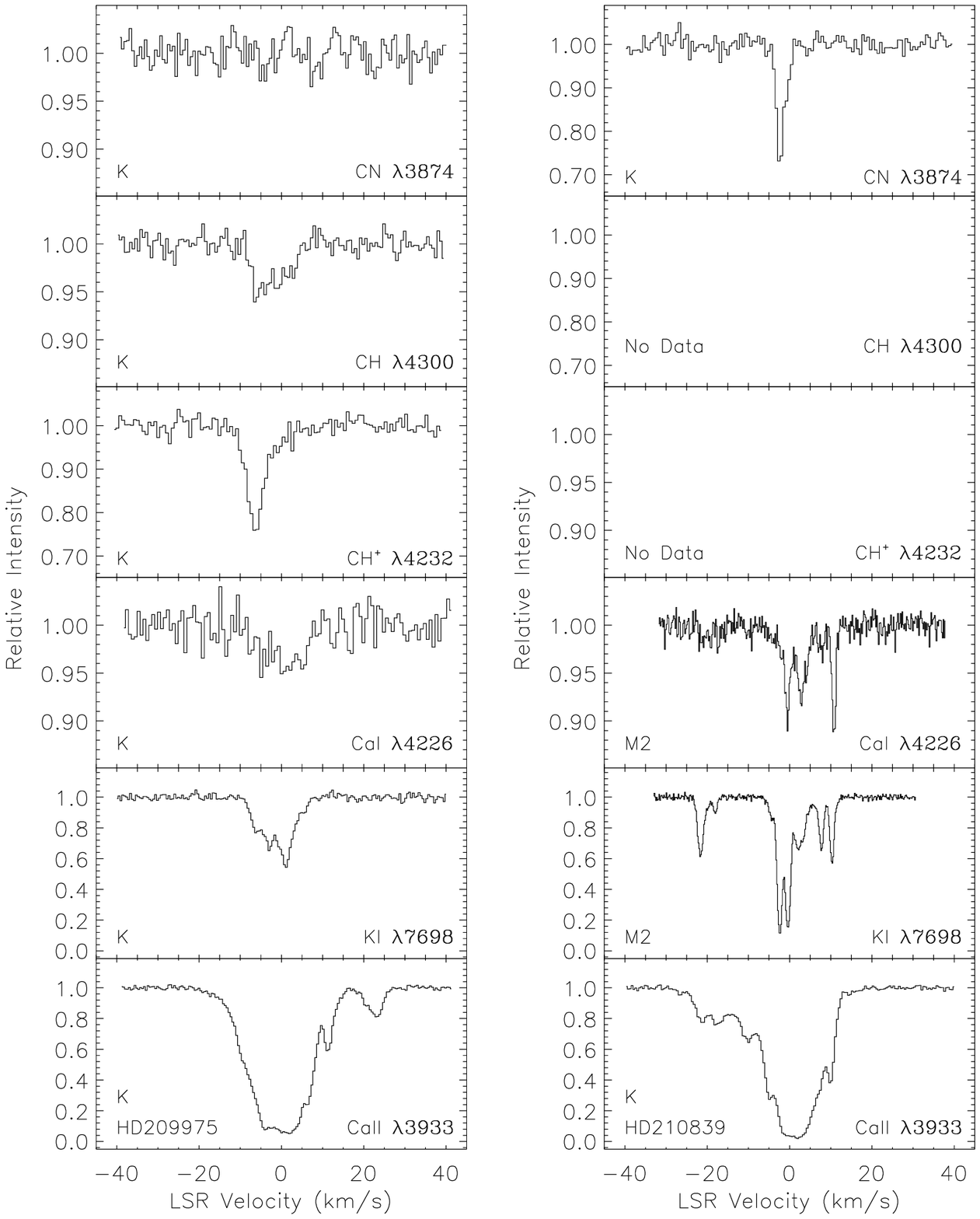} 
\caption{Interstellar  CN, CH, CH$^+$, \ion{Ca}{1}, \ion{K}{1}, and 
  \ion{Ca}{2} absorption profiles toward HD209975 
  and HD210839 (as for Figure 1).} 
\end{figure} 
 
\clearpage 
 
\begin{figure} 
\plotone{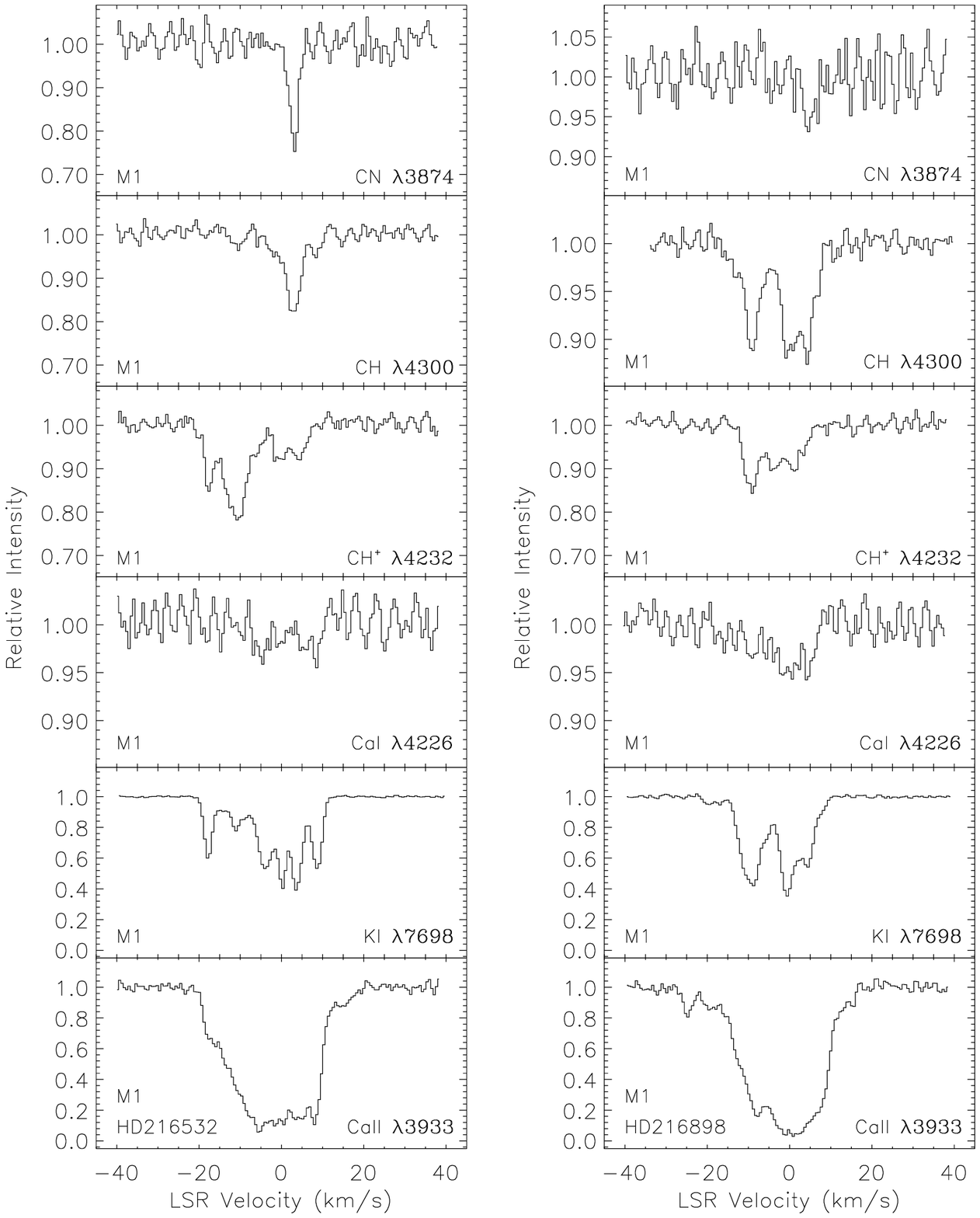} 
\caption{Interstellar  CN, CH, CH$^+$, \ion{Ca}{1}, \ion{K}{1}, and 
  \ion{Ca}{2} absorption profiles toward HD216532 
  and HD216898 (as for Figure 1).} 
\end{figure} 
 
\clearpage 
 
\begin{figure} 
\plotone{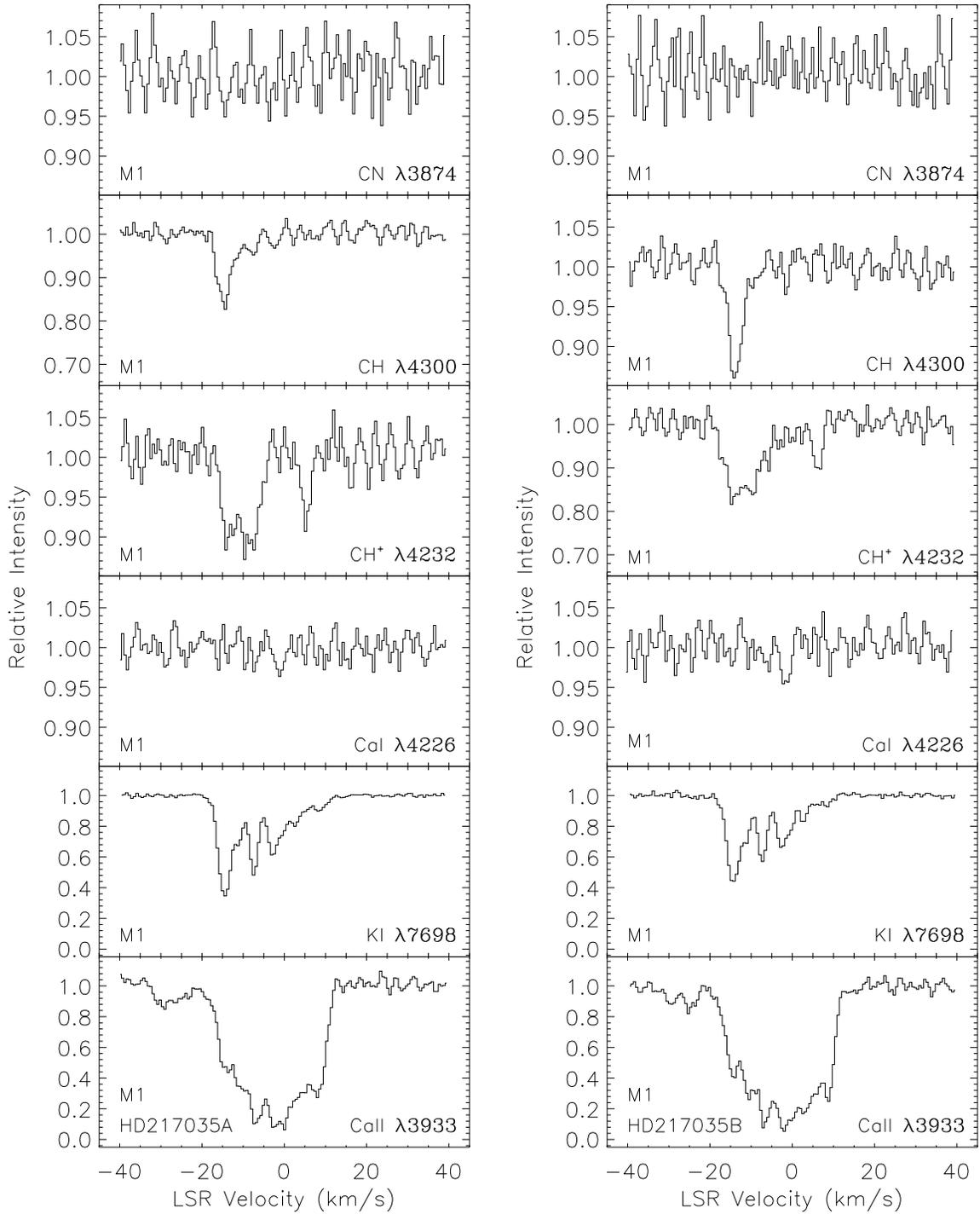} 
\caption{Interstellar  CN, CH, CH$^+$, \ion{Ca}{1}, \ion{K}{1}, and 
  \ion{Ca}{2} absorption profiles toward HD217035A 
  and HD217035B (as for Figure 1).} 
\end{figure} 
 
\clearpage 
 
\begin{figure} 
\plotone{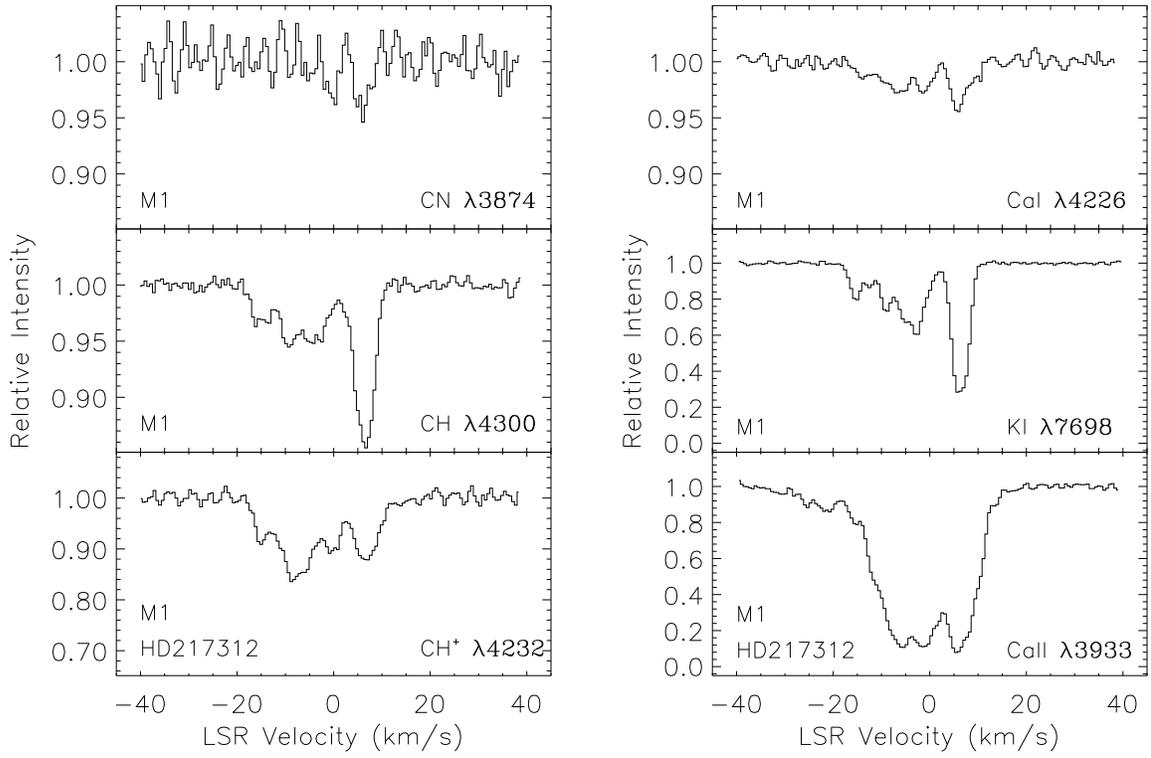} 
\caption{Interstellar  CN, CH, CH$^+$, \ion{Ca}{1}, \ion{K}{1}, and 
  \ion{Ca}{2} absorption profiles toward HD217312 
  (as for Figure 1).} 
\end{figure} 
 
\end{document}